\documentclass{svjour3}                     
\smartqed  
\usepackage[utf8]{inputenc}
\usepackage{graphicx}
\usepackage{comment}
\usepackage{enumerate}
\usepackage{amsmath}
\usepackage{amssymb}
\usepackage{mathrsfs}
\usepackage{lineno, comment, url}
\usepackage[usenames,dvipsnames,svgnames,table]{xcolor}

\usepackage[bbgreekl]{mathbbol}
\usepackage{units}

\usepackage[english]{babel}

\DeclareMathOperator{\divergence}{div}
\DeclareMathOperator{\Tr}{Tr}
\DeclareMathOperator{\stat}{stat}

\DeclareMathOperator{\conv}{conv}
\newcommand{\R}{\ensuremath{{\mathbb R}}}
\newcommand{\uu}{\mathbf{u}}
\newcommand{\rr}{\mathbf{r}}
\newcommand{\xx}{\mathbf{x}}
\newcommand{\eps}{\varepsilon}
\newcommand{\pd}[2]{\ensuremath{\frac{\partial {#1}}{\partial {#2}}}}
\newcommand{\diff}{\mathrm{d}}
\newcommand{\dd}[2]{\ensuremath{\frac{\diff {#1}}{\diff {#2}}}}
\newcommand{\ddd}[2]{\ensuremath{\frac{\diff^2 {#1}}{\diff {#2}^2}}}
\renewcommand{\vec}[1]{\ensuremath{\mathbf{#1}}}
\newcommand{\tensorq}[1]{\ensuremath{\mathbb{#1}}}      
\newcommand{\tensorc}[1]{\ensuremath{\mathrm{#1}}}      
\newcommand{\transpose}[1]{#1^\top}
\newcommand{\identity}{\ensuremath{\tensorq{I}}}
\newcommand{\vecv}{\ensuremath{\vec{v}}}
\newcommand{\vecu}{\ensuremath{\vec{u}}}
\newcommand{\gradv}{\ensuremath{\nabla \vecv}}
\newcommand{\gradsym}{\ensuremath{\tensorq{D}}}
\newcommand{\cstress}{{\tensorq{T}}}
\newcommand{\bydefinition}{\mathrm{def}}
\newcommand{\traceless}[1]{{#1}_{\delta}}
\newcommand{\dcstress}{\cstress - \left( \frac{1}{3}\Tr \cstress \right) \identity}
\newcommand{\dcstresssymb}{{\traceless{\cstress}}}
\newcommand{\absnorm}[1]{\ensuremath{\left|#1\right|}}
\newcommand{\tensorf}[1]{{\mathfrak{#1}}}
\newcommand{\vv}{\vecv}
\newcommand{\cc}{\mathbf{c}}

\newcommand{\JJ}{\dcstresssymb}
\newcommand{\XX}{\gradsym}
\newcommand{\JJconj}{\cstress_\delta^*}
\newcommand{\JJconjconj}{\cstress_\delta^{**}}
\newcommand{\tildeJJ}{\tilde{\cstress}_\delta}

\newcommand{\pp}{\mathbf{p}}
\journalname{Continuum Mechanics and Thermodynamics}

\begin{document}

\title{Non-convex dissipation potentials in multiscale non-equilibrium thermodynamics}
\author{Adam Jane\v{c}ka \and Michal Pavelka$^*$}
\institute{A. Jane\v{c}ka \at
  Mathematical Institute, Faculty of Mathematics and Physics, Charles University in Prague, Sokolovská 83, 186 75 Prague, Czech Republic\\
  \email{janecka@karlin.mff.cuni.cz}           
  \and
  M. Pavelka \at
  Mathematical Institute, Faculty of Mathematics and Physics, Charles University in Prague, Sokolovská 83, 186 75 Prague, Czech Republic\\
  \email{janecka@karlin.mff.cuni.cz}           
  Tel.: +420 22191 3248\\
  \emph{* Corresponding author.}
}

\date{Received: date / Accepted: date}

\maketitle

\begin{abstract}
Reformulating constitutive relation in terms of gradient dynamics (being derivative of a dissipation potential) brings additional information on stability, metastability and instability of the dynamics with respect to perturbations of the constitutive relation, called CR-stability. CR-instability is connected to the loss of convexity of the dissipation potential, which makes the Legendre-conjugate dissipation potential multivalued  and causes dissipative phase transitions, that are not induced by non-convexity of free energy, but by non-convexity of the dissipation potential. CR-stability of the constitutive relation with respect to perturbations is then manifested by constructing evolution equations for the perturbations in a thermodynamically sound way. As a result, interesting experimental observations of behavior of complex fluids under shear flow and supercritical boiling curve can be explained.
 \keywords{Gradient dynamics \and non-Newtonian fluids \and non-convex dissipation potential \and Legendre transformation \and stability \and non-equilibrium thermodynamics}
 \PACS{05.70.Ln \and 05.90.+m}
 \subclass{31C45 \and 35Q56 \and 70S05}
\end{abstract}


\section{Introduction}
\label{sec:introduction}

\begin{figure}[!htbp]
  \centering
  \includegraphics[width=0.6\textwidth]{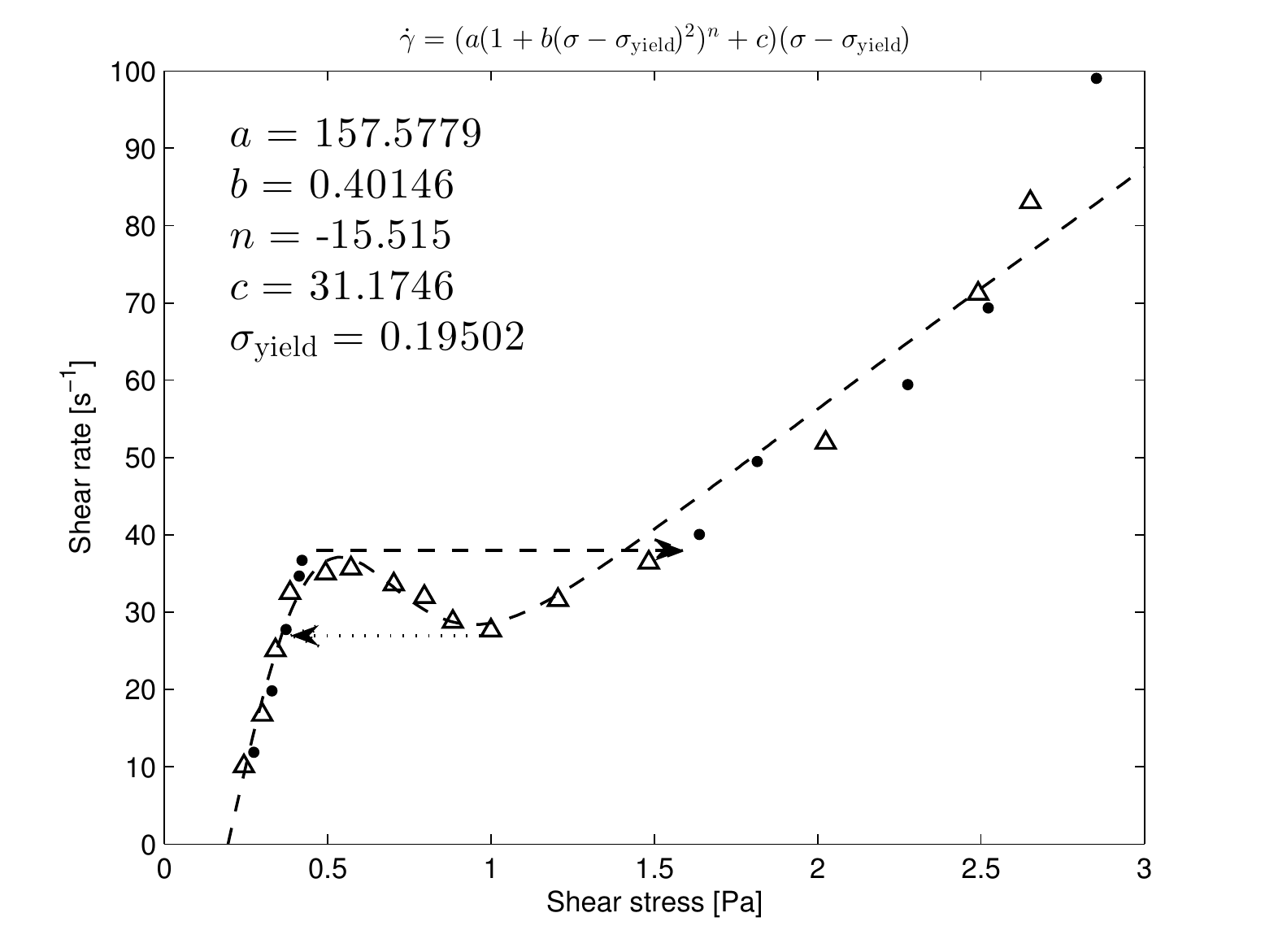}
  \caption{Steady-state stress/shear--rate behavior for a 7.5/7.5 mM TTAA/NaSal solution from the controlled shear stress ($\Delta$) and controlled shear--rate ($\bullet$) experiments by Boltenhagen et al.~\cite{boltenhagen.p.hu.y.ea:observation}. The fit (-\,-) was obtained from the constant applied stress data using one-dimensional form of the constitutive relation~\eqref{eq:le-roux-rajagopal}. \textbf{When increasing shear stress, shear rate varies continuously} (although not monotonically). \textbf{When increasing shear rate, shear stress exhibits a jump.} The dashed arrow indicates the experimentally observed jump in shear stress while the dotted arrow the predicted (see Sec.~\ref{sec.stability}) jump when lowering shear rates (hysteresis). Note that the vertical axis expresses the strain rate, in contrast to the usual habit, and the S-curve in this figure is thus referred to as vertical.}
  \label{fig.exp}
\end{figure}

Consider two concentric cylinders, the outer rotating with respect to the inner one. The gap between the cylinders is filled with a non-Newtonian fluid. In such an experiment either the force required to rotate the outer cylinder (shear stress) or speed of the rotation (shear rate) can be controlled, the other being measured, see Fig.~\ref{fig.exp}.\footnote{We assume that the measured shear stress and shear rate corresponds to the actual properties of the fluid within the gap.} An interesting experimental observation is that when varying shear stress, shear rate behaves continuously, while when varying shear rate, shear stress exhibits a jump \cite{boltenhagen.p.hu.y.ea:observation}.

We ask the following two questions: 
\begin{enumerate}
\item How to describe the relation between shear stress and shear rate by means of a constitutive relation compatible with non-equilibrium thermodynamics?
\item What is the reason that when controlling the shear stress, the response (shear rate) is continuous, while when controlling the shear rate, the response (shear stress) is discontinuous?
\end{enumerate}

The first question is answered in Sec.~\ref{sec.CR}, where an appropriate model is fitted to the experimental data and then reformulated in terms of gradient dynamics. The second question is partially answered in Sec.~\ref{sec.conjugate}, where the multivalued conjugate gradient structure is identified. Subsequently, in Sec.~\ref{sec.CR-thermodynamics}, the dynamics is lifted to an extended state space. This enables us to discuss various regimes of the dynamics and its stability, metastability and instability with respect to perturbations of the constitutive relation---the CR-stability, and to complete answering the second question in Sec.~\ref{sec.stability}, which is the essential result of this paper. Moreover, we propose that the phenomenon of critical heat flux in Sec. \ref{sec.CHF} could be approached by the method of non-convex dissipation potentials.

Novelty of this paper lies in the following points. A dissipation potential is identified that yields constitutive relation~\eqref{eq:le-roux-rajagopal}. The dissipation potential is not convex everywhere (it is convex in the vicinity of equilibrium), and CR-stability of the dynamics generated by this dissipation potential (with respect to perturbations of the dissipation potential) is discussed in the context of mesoscopic multiscale thermodynamics. The conjugate dissipation potential, which is obtained by a form of Legendre transformation, is multivalued, which, to our best knowledge, is mentioned in the thermodynamic literature for the first time whatsoever. Moreover, a thermodynamic reason is identified which favors seeking constitutive relation in the form $\gradsym(\dcstresssymb)$, i.e. shear rate as a function of shear stress. Finally, the experimental behavior presented in Fig.~\ref{fig.exp} is explained. A model based on a non-convex dissipation potential is also proposed for the critical heat flux in Sec \ref{sec.CHF}.

\section{Constitutive relation}
\label{sec.CR}

The experimental dependence of shear rate on shear stress plotted in Fig.~\ref{fig.exp} can be described by a model of non-Newtonian fluid introduced by Le~Roux and Rajagopal~\cite{roux.c.rajagopal.kr:shear}
\begin{equation}
  \label{eq:le-roux-rajagopal}
  \gradsym = \left[ \alpha \left( 1 + \beta \absnorm{\dcstresssymb}^2 \right)^n + \gamma \right] \dcstresssymb,
\end{equation}
where $\alpha$ and $\beta$ are positive numbers, $\gamma$ is a non-negative number, $n$ is a real number, $\gradsym =_\bydefinition \frac{1}{2} \left( \gradv + \transpose{\gradv} \right)$ is the symmetric part of the velocity gradient and $\dcstresssymb =_\bydefinition \dcstress$ is the deviatoric part of the Cauchy stress $\cstress$. Setting $\gamma = 0$, constitutive relation~\eqref{eq:le-roux-rajagopal} reduces to the stress power-law model studied by Málek et al.~\cite{malek.j.prusa.ea:generalizations}.  The fit is plotted in Fig.~\ref{fig.exp}. 

Note that only the deviatoric part of the stress tensor is taken into account and that the fluid is considered to be effectively incompressible. It is possible that abandoning such a constraint could lead to other interesting behavior, but the purpose of this paper is achieved while respecting this constraint.

Constitutive relation \eqref{eq:le-roux-rajagopal} can be seen as a special case of general implicit constitutive relation \cite{rajagopal.kr:on*3,rajagopal.kr:on*4}
\begin{equation}
  \label{eq:general-implicit-relation}
  \tensorf{h} (\dcstresssymb, \gradsym) = \tensorq{0},
\end{equation}
where $\tensorf{h}$ is a tensorial function. It is however not clear whether the fully implicit case brings any advantage and whether it is physically substantiated. On the other hand, the special case $\gradsym(\dcstresssymb)$ has a sound thermodynamic meaning and practical advantages, see Sec.~\ref{sec.CR-thermodynamics} and \cite{perlacova.t.prusa.v:tensorial}.

\begin{figure}[!htbp]
  \centering
  \includegraphics[width=0.6\textwidth]{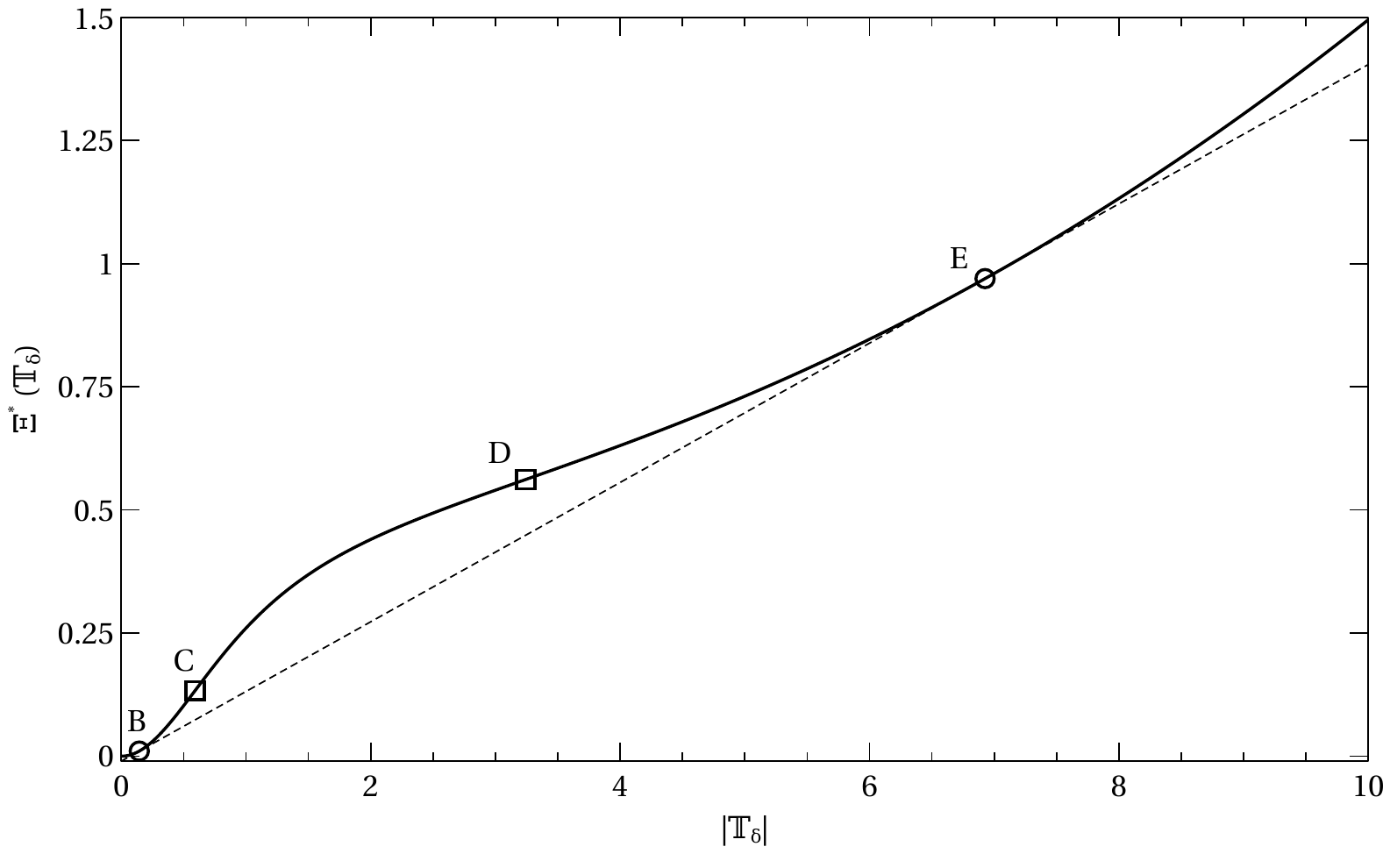}
  \caption{Dissipation potential $\Xi^{*} \left( \dcstresssymb \right)$ corresponding to the constitutive relation~\eqref{eq:le-roux-rajagopal} with parameter values $\alpha = 1$, $\beta = 1$, $\gamma = 0.02$ and $n = -2$. The segment $C-D$ is not convex, thereby the dissipation potential is not CR-stable in this region. The dashed tangent line $B-E$ indicates the convex hull of the potential and coincides with the Maxwell lever rule construction. The segments $B-C$ and $D-E$ correspond to the CR-metastable regions (compare with Fig.~\ref{fig:flow-curve}).}
  \label{fig:dissipation-potential}
\end{figure}

Constitutive relation \eqref{eq:le-roux-rajagopal} can be also regarded as a consequence of gradient dynamics\footnote{Gradient dynamics is a cornerstone of the GENERIC framework \cite{GO,OG}, where it generates the dissipative evolution. It has strong connection to the principle of large deviations \cite{Mielke-Peletier}, and it can be related to other techniques associated with the maximization of entropy production \cite{SEA,Rajagopal2004}, where for instance quantum dissipation is addressed. However, we prefer using gradient dynamics, since it seems to be more straightforward in the case of non-quadratic dissipation potentials \cite{janecka.a.pavelka.m:gradient}. Note that an alternative non-potential version of GENERIC is preferred by others \cite{HCO,Hutter2013}. Although the latter version is more general as it allows for antisymmetric dissipative coupling, it is not clear whether such a coupling is necessary and correct \cite{Miroslav-WhyGeneric}. We prefer the potential version of GENERIC because reversibility and irreversibility are then clearly distinguished \cite{PRE2014} and because we can then perform Legendre transformations, which are crucial for the manuscript at hand.}
\begin{subequations}
  \begin{equation}
    \label{eq.gradient}
    \gradsym = \pd{\Xi^*}{\dcstresssymb},
  \end{equation}
  where the shear rate is a derivative of the dissipation potential $\Xi^*$ with respect to the shear stress.\footnote{The derivative is to be interpreted as the functional derivative in general. It will be also denoted simply by a subscript, i.e. $\Xi^*_{\dcstresssymb}$.} Alternatively, one can use the conjugate form
  \begin{equation}
    \label{eq.cgradient}
    \dcstresssymb = \pd{\Xi}{\gradsym},
  \end{equation}
\end{subequations}
where the dissipation potential $\Xi$ depends on the shear rate, and it is conjugate to $\Xi^*$ by means of the Legendre transformation, see Sec.~\ref{sec.conjugate} for more details. The former case, Eq.~\eqref{eq.gradient}, is covered by dissipation potential
\begin{equation}
  \label{eq.theta}
  \Xi^* \left( \dcstresssymb \right) = 
  \begin{cases}
    \frac{\alpha}{2 \beta (n + 1)} \left[ (1 + \beta \absnorm{\dcstresssymb}^2)^{n+1} - 1 \right] + \frac{\gamma}{2} \absnorm{\dcstresssymb}^2, & n \neq -1, \\
    \frac{\alpha}{2 \beta} \ln \left(1 + \beta \absnorm{\dcstresssymb}^2 \right) + \frac{\gamma}{2} \absnorm{\dcstresssymb}^2, & n = -1,
  \end{cases}
\end{equation}
whose derivative gives Eq.~\eqref{eq:le-roux-rajagopal} in the sense of Eq.~\eqref{eq.gradient}. This dissipation potential is plotted in Fig.~\ref{fig:dissipation-potential}. The latter case, Eq.~\eqref{eq.cgradient}, is covered by the dissipation potential $\Xi$ specified in Sec.~\ref{sec.conjugate}. 

The important feature of the dissipation potential $\Xi^*$, Eq.~\eqref{eq.theta}, is that it is convex in the vicinity of zero while non-convex for higher shear stresses. Convexity is then regained for even higher shear stresses. By analogy with equilibrium thermodynamics \cite{Callen}, it can be expected that the convex parts generate CR-stable or CR-metastable evolution while the non-convex parts lead to CR-instability. This is indeed so as discussed in Sec.~\ref{sec.CR-thermodynamics} and Sec.~\ref{sec.stability}. 

\subsection{Second law of thermodynamics}
\textbf{On the second law of thermodynamics and convexity}: Assume for a moment that the dissipation potential $\Xi(\XX)$ is smooth and convex. The potential is zero at the origin an thus is positive everywhere except in the thermodynamic equilibrium. Due to the convexity, the origin is also the only point where derivative of the dissipation potential is zero, $\JJ = \Xi_{\XX}|_{\XX=0} = \tensorq{0}$. 
The conjugate dissipation potential $\Xi^*(\JJ)$, which is formed by Legendre transformation, is also convex and zero at the origin. Moreover, it holds that
\begin{equation}
\langle \XX, \JJ\rangle = \underbrace{\Xi(\XX)}_{\geq 0} + \underbrace{\Xi^*(\JJ)}_{\geq 0}\geq 0,
\end{equation}
see e.g. p. 267 in \cite{Roubicek}. Entropy production is the product of thermodynamic forces and fluxes as usually in  \cite{dGM} or in the GENERIC evolution equation for momentum \eqref{eq.evo.uu.GENERIC}, and this last inequality represents the positive sign of entropy production. Convexity of the dissipation potential thus implies compatibility with the second law of thermodynamics.

\textbf{On the second law and non-convexity}: Convexity of the dissipation potential is a sufficient condition for compatibility with the second law. However, it is not a necessary condition. The second law is fulfilled if and only if the dissipation potential generates non-negative entropy production, 
\begin{equation}\label{eq.2nd}
  \langle \XX, \Xi_\XX\rangle \geq 0 \qquad\mbox{or}\qquad \langle \JJ, \Xi^*_\JJ\rangle \geq 0.
\end{equation} 
Practically always in non-equilibrium thermodynamics this condition is required to hold locally (not only after integration over the whole system). Moreover, let us consider only the case where the dissipation potential depends only on a norm of the thermodynamic force or flux (as always in this paper). The local version of condition \eqref{eq.2nd} is then
\begin{equation}
  0 \leq \JJ : \Xi^*_\JJ = \absnorm{\JJ} \frac{\JJ}{\absnorm{\JJ}} : \Xi^*_{\absnorm{\JJ}} \pd{\absnorm{\JJ}}{\JJ} = 
  \absnorm{\JJ} \frac{\JJ}{\absnorm{\JJ}} : \Xi^*_{\absnorm{\JJ}} \frac{\JJ}{\absnorm{\JJ}}  
  = \absnorm{\JJ} \Xi^*_{\absnorm{\JJ}},
\end{equation}
which is satisfied for instance if the function $\Xi^*(|\JJ|)$ is monotonous, see Fig.~\ref{fig:dissipation-potential}. In this sense, a non-convex dissipation potential which is monotonous as a function of $\absnorm{\JJ}$ fulfills the second law of thermodynamics.

A generalization to dissipation potentials that are not necessarily simple functions of a norm of the forces or fluxes was found for instance in \cite{Mira-Milan} (Eqs. (28)-(30) therein). 


It should be also borne in mind that condition \eqref{eq.2nd} expresses non-negativity of entropy production only if the thermodynamic potential whose derivatives generate the thermodynamic forces is convex. Therefore, we always assume that entropy is a concave function of the state variables and that the thermodynamic potential is convex.

We believe that the experimental observations can be understood also without non-convex dissipation potentials, but on more detailed levels of description. The non-convex dissipation potential should be a result of a reduction from a more detailed level of description equipped with a convex dissipation potential. It is not clear, however, what is the more detailed level of description. Perhaps a further extension in the sense of internal variables (that can be often seen as a CR-extension) could generate the same results using a convex dissipation potential. But unless such extension is known, let us proceed with a non-convex dissipation potential.

\subsection{Approach to equilibrium and existence}
Convexity of the dissipation potential near equilibrium, which can be interpreted as approximately quadratic behavior, is moreover in tight relation with stability of the equilibrium state. Apart from the convexity of the dissipation potential, convexity of the energy plays a crucial role in stability analysis. When these two requirements are satisfied, the equilibrium state is stable with respect to small fluctuations, as can be seen for example from analysis of the spectrum of the linearized GENERIC near equilibrium, see~\cite{PRE2014}, or as discussed in the paper by Matolcsi~\cite{Matolcsi-Lyapunov}. Thermodynamic equilibrium is stable if the dissipation potential is convex and the entropy concave near equilibrium.

On the other hand, although convexity of the dissipation potential (also monotonicity of the derivative) implies increase of the entropy by condition \eqref{eq.2nd}, it does not make up for an exact mathematical proof of the existence of solutions and of the approach to equilibrium from arbitrarily far-from-equilibrium states. It is, however, an important and essential ingredient from both the physical and the mathematical points of view. This can be well illustrated on the case of the Boltzmann equation where the exact mathematics has been done.

Boltzmann equation is an illustration of GENERIC with the non-quadratic but convex dissipation potential, see~\cite{GO}. The proof of the global existence of solutions has been done by Di Perna and Lions~\cite{Lions}, for which the latter author received the Fields Medal. The Boltzmann $H$-theorem is not sufficient but plays an essential role in the proof. The approach to the equilibrium state was also shown by Desvilletes and Villani \cite{Desvilettes-Villani}, where again the Boltzmann $H$-theorem does not suffice but plays an essential role.

Summing up, the increase of the entropy as implied by monotonicity of the dissipation potential does not, indeed, imply immediately all the mathematical rigor in the proofs of the existence of solutions and the approach to the equilibrium from any far-from-equilibrium state, but it is an essential ingredient in such proofs (that have been done only for a few mesoscopic time evolution equations so far).


\subsection{Shear and vorticity banding}

Let us now return to Fig. \ref{fig.exp}. It is possible to interpret the $\gradsym(\dcstresssymb)$ dependence in Fig.~\ref{fig.exp} (open triangles) in terms of gradient dynamics. For each $\dcstresssymb$ one has a unique value of $\gradsym$, which is the slope of the function $\Xi^*$ at the particular value of $\dcstresssymb$. In the next section, we try to answer how to interpret the dependence $\dcstresssymb(\gradsym)$.

  Before going on with the details on the conjugate dissipation potential, let us briefly comment on the relation of this manuscript to the gradient and vorticity banding phenomena, excellently reviewed in~\cite{Fielding,Olmsted}. Shear banding is usually experimentally observed in the cylindrical Couette flow (as in this paper), where there might develop instabilities in form of (visible) bands. If the bands develop in the radial direction, the phenomenon is referred to as gradient banding while if they develop in the direction of the rotational axis, the phenomenon is referred to as vorticity banding. 

  Gradient banding usually leads to stress--strain curves with multiple shear rates corresponding to a single shear stress (''horizontal'' S-curve) while vorticity banding leads to curves where multiple shear stresses correspond to a single shear rate (''vertical'' S-curve). Although such distinction between the two phenomena is typical, it is not always true \cite{Olmsted}. When crossing the region of unstable shear rates in the case of gradient banding, the region of higher shear rates is usually characterized by lower effective viscosity, i.e. the fluid is shear thinning. On the other hand, vorticity banding often corresponds to increased viscosity---shear thickening fluids. However, there have been reported experiments where such distinction was not valid \cite{Olmsted}. 

  Since it is difficult to characterize gradient/vorticity banding in the terms of horizontal/vertical S-curves or shear thinning/thickening fluids, we shall discuss the S-curves themselves. There are several models leading to horizontal S-curves for the stress--strain dependency (multiple shear rates for a single shear stress). For instance, the Johnson--Segalman model \cite{Johnson-Segalman} or rather its weakly non-local variant \cite{Olmsted-Radulescu}, thermodynamics with internal variables \cite{Verhas,Asszonyi2015}, viscoelastic models undergoing scission of polymeric chains \cite{Miroslav-micelles} or viscoelastic models with relaxation of internal structure \cite{Miroslav-suspensions,Germann-Beris} indicate the horizontal S-shape of the stress--strain relation. The literature is rather vast, and we do not attempt to make the list of references complete. Interested reader could consult, for example, recent special issue of the Journal of Rheology \cite{Fielding-special}. Most of the models usually exhibit shear thinning, and it is believed that shear thickening should be captured by models with non-linear dependence of the constitutive relation on the shear stress \cite{Radulescu}, which is a motivation for the constitutive relation~\eqref{eq:le-roux-rajagopal}. This paper thus aims at modeling of the vertical S-shaped stress--strain curve with shear thickening, typically observed in vorticity banding.

\section{Conjugate representation}
\label{sec.conjugate}

The purpose of this section is to present details on the Legendre duality between $\Xi$ and $\Xi^*$. A consequence of the non-convexity of $\Xi^*$ is that $\Xi$ is multivalued, i.e. there are multiple shear stresses corresponding to certain shear rates. The shear rates that correspond to multiple stresses then constitute the CR-metastable and CR-unstable regions of the dynamics, which are discussed in Sec.~\ref{sec.stability}. In fact, Sec.~\ref{sec.stability} contains the essence of this paper, and reading it before the following more technical parts could be advisable. Note also that Legendre transformation can be regarded as constrained extremization of the generating potential, which often turns out to be a practical method, see e.g. \cite{Junker2014}. 

The balance of linear momentum reads
\begin{equation}
  \label{eq.u.evo}
  \pd{\uu}{t} = \underbrace{-\nabla p - \divergence\left(\frac{\uu\otimes\uu}{\rho}\right)}_{=\mbox{reversible terms}} + \divergence(\dcstresssymb),
\end{equation}
where $\uu$ is the density of momentum (per volume). The irreversible part of this evolution equation is in the form of divergence of the shear stress $\dcstresssymb$ (thermodynamic flux). Shear rate $\gradsym$ is thus interpreted as a thermodynamic force. The reversible terms, see Appendix \ref{sec.rev} for the definition of reversibility and irreversibility, are the gradient of the pressure and the divergence of convective terms, and their Hamiltonian nature is discussed in App.~\ref{sec.hydro}.

\begin{figure}[!htbp]
  \centering
  \includegraphics[width=0.6\textwidth]{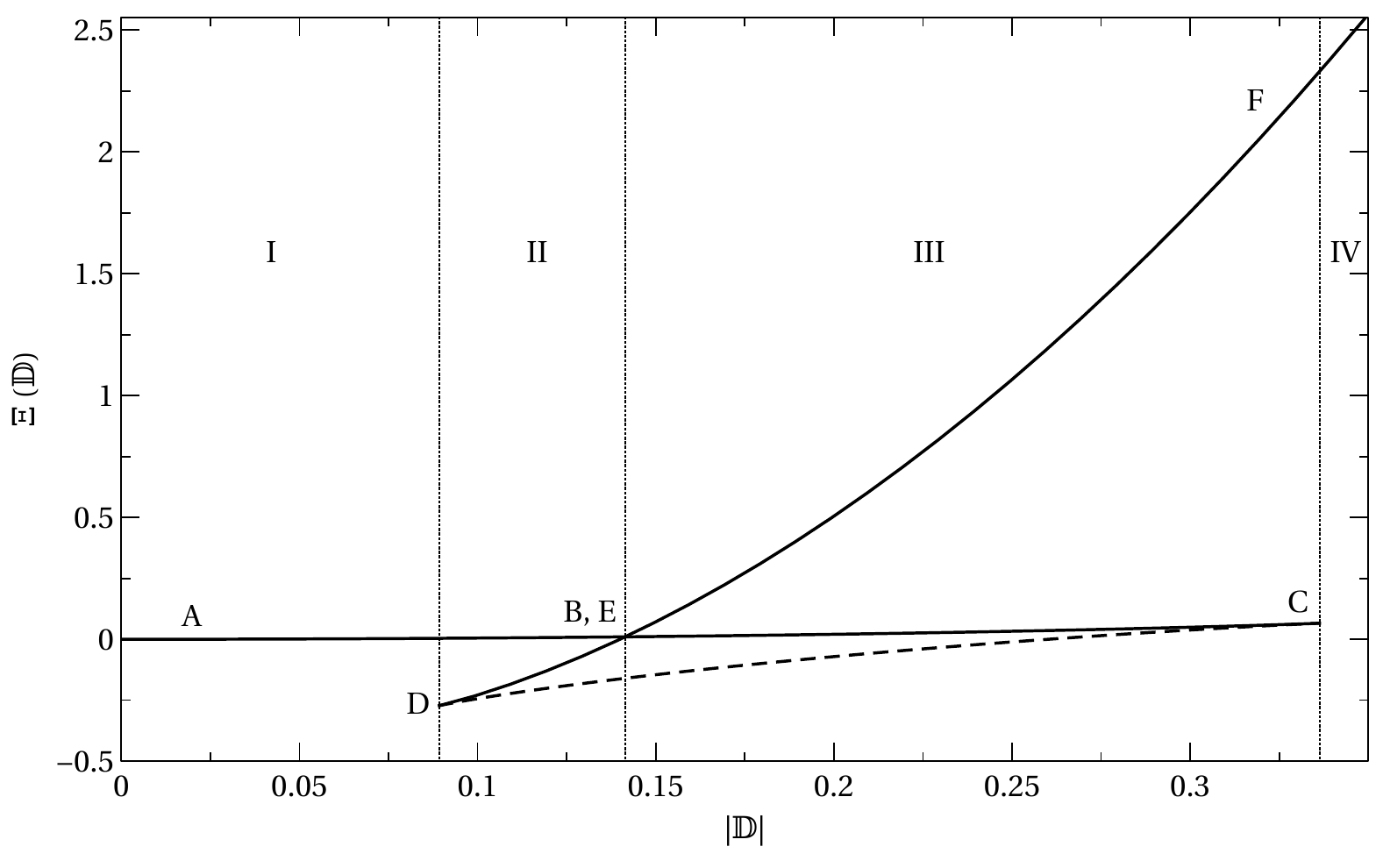}
  \caption{Conjugate dissipation potential $\Xi \left( \gradsym \right)$ corresponding to the constitutive relation~\eqref{eq:le-roux-rajagopal} is obtained from the dissipation potential $\Xi^* \left( \dcstresssymb \right)$ (Fig.~\ref{fig:dissipation-potential}) using the generalized Legendre transformation~\eqref{eq.Xi}. In regions II and III, the potential is multivalued. The dashed line $C-D$ is the CR-unstable branch, lines $B-C$ and $D-E$ are CR-metastable branches, and the curve $A-B,E-F$ is globally CR-stable and corresponds to the convex hull of the dissipation potential $\Xi^*$. See Sec.~\ref{sec.stability} for detailed explanation.}
  \label{fig:conjugate-dissipation-potential}
\end{figure}

Only the field of momentum density plays the role of state variable, since we assume that the fluid is incompressible ($\divergence \vecv = 0$) and isothermal. When carrying out the CR-extension below (Sec. \ref{sec.promotion}), also the field of conjugate stress tensor becomes a state variable.

\begin{subequations}\label{eq.LT.Xic.Xi}
Equation \eqref{eq.gradient} can be then rewritten as 
\begin{equation}
  \label{eq.Xic.JJ}
  \left. \Xi^*_\JJ \right|_{\tildeJJ(\XX)} = \XX,
\end{equation}
and can be regarded as an equation for $\JJ$. Denoting the solutions to this equation by $\tildeJJ (\XX)$, the equation can be further interpreted as 
\begin{equation}
  \label{eq.Xic.Xi}
  \pd{}{\JJ} \left( -\Xi^*(\JJ) + \langle \JJ, \XX \rangle \right) \Big|_{\tildeJJ (\XX)} = \tensorq{0},
\end{equation}
where $\langle\bullet,\bullet\rangle$ is the $L^2$ scalar product---spatial integral of product of the two arguments. The last equation leads to the formulation of the dissipation potential $\Xi$ conjugate to $\Xi^*$ 
\begin{equation}
  \label{eq.Xi}
  \Xi(\XX) =_\bydefinition \left( -\Xi^*(\JJ) + \langle \JJ,\XX \rangle \right) \Big|_{\tildeJJ (\XX)}.
\end{equation}
Note that we do not require the solution to be unique, which means that we do not interpret $\tildeJJ (\XX)$ as a function, but rather as a graph in the $(\XX,\JJ)$ plane or as a collection of functions. Therefore, dissipation potential $\Xi$ is multivalued as can be seen from Fig.~\ref{fig:conjugate-dissipation-potential}. Although the dissipation potential is multivalued, it is uniquely determined and the generalized (multivalued) Legendre transformation is invertible\footnote{Invertible in the sense that the backward Legendre transformation restores the original function (or rather graph).} \cite{dorst.l.boomgaard.r:analytical,dorst.l.boomgaard.r:morphological}.
\end{subequations}

\begin{figure}[!htbp]
  \centering
  \includegraphics[width=0.6\textwidth]{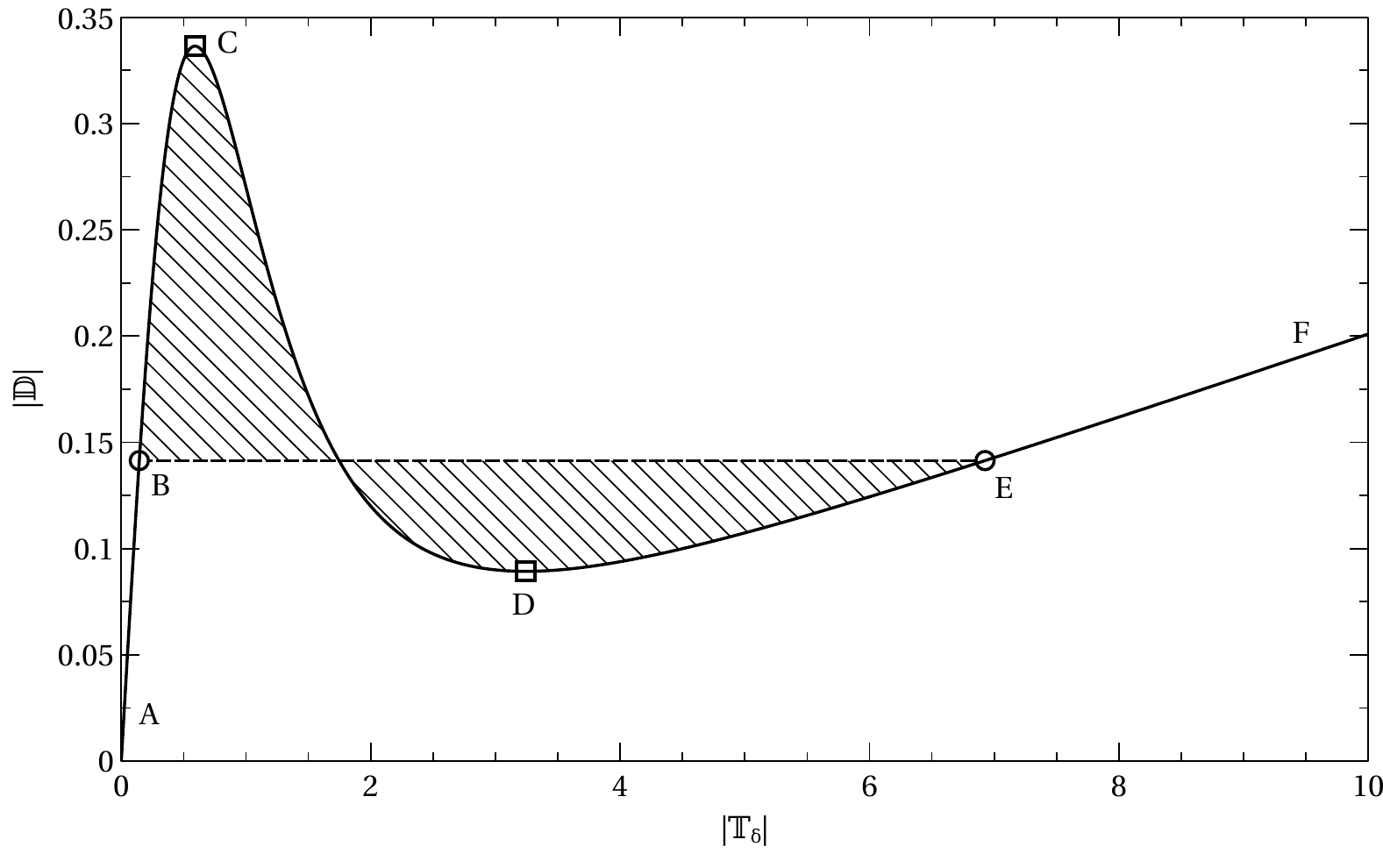}
  \caption{Flow curve (the constitutive relation) given by Eq.~\eqref{eq:le-roux-rajagopal} with parameter values $\alpha = 1$, $\beta = 1$, $\gamma = 0.02$ and $n = -2$. The line $B-E$ is constructed using the Maxwell lever rule, so that the shaded areas are equal. Since the segment $C-D$ corresponds to the region where the dissipation potential $\Xi^*$ loses convexity, it is CR-unstable, as commented in Sec.~\ref{sec.stability}. The lines $B-C$ and $D-E$ are CR-metastable---they coincide with the lower CR-stable branches of the dissipation potential $\Xi$, see Fig~\ref{fig:conjugate-dissipation-potential}. Note that the vertical axis expresses the strain rate, in contrast to the usual habit, and the S-curve in this figure is thus referred to as a vertical. }
  \label{fig:flow-curve}
\end{figure}

\begin{subequations}\label{eq.LT.Xi.Xic}
Using Eq.~\eqref{eq.Xic.JJ}, we obtain from Eq.~\eqref{eq.Xi} that
\begin{equation}
  \label{eq.Xi.XX}
  \Xi_{\XX} = \tildeJJ (\XX),
\end{equation}
solution of which are denoted by $\tilde{\XX}(\JJ)$. These solutions can be regarded as the same graph in the $(\XX,\JJ)$ plane as solutions $\tildeJJ (\XX)$. An example of such a graph, which expresses simply the constitutive relation is plotted in Fig.~\ref{fig:flow-curve}.
Equation~\eqref{eq.Xi.XX} can be rewritten as
\begin{equation}
  \pd{}{\XX} \left(-\Xi(\XX) + \langle\XX,\JJ\rangle\right)\Big|_{\tilde{\XX}(\JJ)} = \tensorq{0},
\end{equation}
 which finally leads back to the original dissipation potential,
\begin{equation}
  \Xi^*(\JJ) = -\Xi \left(\tilde{\XX}(\JJ)\right) + \left\langle\tilde{\XX}(\JJ), \JJ\right\rangle,
\end{equation}
derivative of which gives Eq.~\eqref{eq.Xic.JJ}. We have thus performed Legendre transformation from $\Xi^*$ to $\Xi$ and back. Interested reader can consult App.~\ref{sec.Legendres} for other forms of Legendre-like transformations.
\end{subequations}

The graph $\tildeJJ (\XX)$ or $\tilde{\XX}(\JJ)$ is plotted in Fig.~\ref{fig:flow-curve}, where also the lever rule is indicated. Of course, on one hand, we have 
\begin{equation}
  \int_{|\dcstresssymb|(B)}^{|\dcstresssymb|(E)} |\gradsym(T)| \,\diff T = \int_{|\dcstresssymb|(B)}^{|\dcstresssymb|(E)} \Xi^*_{|\dcstresssymb|(T)} \,\diff T = \Xi^*(|\dcstresssymb(E)|)-\Xi^*(|\dcstresssymb(B)|),
\end{equation}
while on the other hand from Fig.~\ref{fig:dissipation-potential} it is clear that
\begin{equation}
  \frac{\Xi^*(\dcstresssymb(E))-\Xi^*(\dcstresssymb(B))}{|\dcstresssymb(E)|-|\dcstresssymb(B)|} = \Xi^*_{|\dcstresssymb(E)|}= \Xi^*_{|\dcstresssymb(B)|}.
\end{equation}
Combining these two results leads to 
\begin{equation}
  \int_{|\dcstresssymb|(B)}^{|\dcstresssymb|(E)} |\gradsym(T)| \,\diff T =(|\dcstresssymb|(E) - |\dcstresssymb|(B))\:\underbrace{|\gradsym(|\dcstresssymb(E)|)|}_{=|\gradsym(|\dcstresssymb|(B)|)|},
\end{equation}
which means that the shaded areas in Fig.~\ref{fig:flow-curve} are equal---the Maxwell lever rule  known from equilibrium thermodynamics \cite{Callen}.
 
The Maxwell lever rule plays an analogical role in equilibrium phase transitions. The hysteresis is observed if fluctuations are sufficiently small. If they are large enough, the system evolves along the straight Maxwell line. The same is true in the case of the non-convex dissipation potential.

The Legendre duality between $\Xi$ as a function of forces and $\Xi^*$ as a function of fluxes is often advantageous because one of the forms can be more accessible. The former choice is used for example in chemical kinetics \cite{Grchemkin} while the latter form for example used in plasticity modeling \cite{Mielke2003}.

\textbf{Stability of constitutive relation}: At many places within this paper, we discuss CR-stability of constitutive relations generated by non-convex dissipation potentials. By CR-stability of constitutive relations we mean whether the constitutive relation is CR-stable with respect to perturbations. If the dissipation potential is convex everywhere, the dual (Legendre transformed) dissipation potential is convex as well, and for each value of the thermodynamic force there is only one value of the flux. On the other hand, if the potential is non-convex, the dual dissipation potential is multi-valued and for some values of thermodynamic forces (or fluxes) there are several possible values of fluxes (or forces). The graph of the constitutive relation then has several branches as in Fig.~\ref{fig:conjugate-dissipation-potential}. In such regions where multiple admissible fluxes (or forces) are present, perturbation of the flux can cause evolution of the system towards another branch of the graph of the constitutive relation. The constitutive relation is said to be CR-unstable if any perturbation leads to selection of a different branch, CR-metastable if only large enough perturbation switches the system to a different branch, and CR-stable if the current branch of the constitutive relation is not changed by any perturbation of the flux.

\section{Extended hydrodynamics (CR-extension)}
\label{sec.CR-thermodynamics}

  In this section the set of hydrodynamic state variables\footnote{The state variables are $(\rho,\uu,s)$ in the energetic representation or $(\rho, \uu, e)$ in the entropic representation.} is expanded by promoting shear stress to an independent state variable, which is the reason for the name extended hydrodynamics. The idea of extending the set of state variables is ubiquitous in non-equilibrium thermodynamics, for example Extended Irreversible Thermodynamics (EIT) \cite{Jou-EIT} or thermodynamics with internal variables \cite{Van-Berezovski}. For simplicity, only the isothermal setting is considered in this paper. The isothermal hydrodynamic evolution is thus enriched to a form of extended hydrodynamics. The extra state variable has its own evolution equation, which will be shown to be the derivative of a \textbf{multiscale thermodynamic Lyapunov} (MTL) function with respect to the conjugate of the extra state variable (Sec.~\ref{sec.fast}). The extra state variable then evolves towards the maxima of the Lyapunov function. The figures in Sec.~\ref{sec.stability} show how the Lyapunov function depends on the shear rate and towards what values the shear stress evolves. This will give a graphical representation of the CR-stability analysis of the system. Motivation for reading this rather technical section can thus be found in Sec.~\ref{sec.stability}.

Dissipation potential $\Xi^*$ is convex near equilibrium and for high shear stresses, but it loses convexity in the intermediate range of the shear stress. Convexity leads to existence of a Lyapunov functional, see e.g. \cite{GO}, and Braun--Le Chatelier principle \cite{Braun-LeChat}, that guarantees stability of the evolution. The former paper is crucial for understanding everything that follows. Instability can be thus anticipated in the non-convex regions. Let us now have a look at a non-equilibrium thermodynamic description of stability, metastability and instability. The analysis is formulated within a recent framework developed by Grmela~\cite{MG-CR}, the CR-thermodynamics (CR stands for constitutive relation), which is a form of mesoscopic multiscale thermodynamics.\footnote{We use a slightly different notation compared with the original paper \cite{MG-CR}. Instead of thermodynamic potentials, we use the negatives of the potentials, and we call them MTL functions. The reason is that we can then always use the same form of Legendre transformation. Another difference is that the extra variable is denoted by $\JJconj$ instead of $\JJ$ so that the usual meaning of flux is kept.}

\subsection{GENERIC structure of hydrodynamics}

Evolution of momentum density is governed by Eq.~\eqref{eq.u.evo}, which can be rewritten as (within the GENERIC framework)
\begin{equation}
  \label{eq.evo.uu.GENERIC}
  \pd{\uu}{t} = \left( \pd{\uu}{t} \right)_{rev} + \left. \Xi_{\uu^*} \right|_{\uu^* = \Psi^{(CH\rightarrow E)}_\uu}.
\end{equation}
The reversible part is Hamiltonian, see in Appendix \ref{sec.hydro}, and it expresses the Euler equations of compressible hydrodynamics. The MTL function (the negative of the thermodynamic potential from \cite{MG-CR}) driving classical hydrodynamics (CH) to equilibrium (E) is 
\begin{equation}
  \Psi^{(CH\rightarrow E)} = \int \diff\rr\, S(\rho,\uu,e) - \frac{1}{T_0} E(\rho,\uu,e) +\frac{\mu_0}{T_0} M(\rho,\uu,e),
\end{equation}
where $e$ is total energy density (per volume).
Total entropy is
\begin{equation}\label{eq.S.hydro}
  S = \int \diff\rr\, s\left(\rho, e - \frac{\uu^2}{2\rho} \right),
\end{equation}
where $s(\rho,\eps)$ denotes local equilibrium entropy density (which depends on density and internal energy density $\eps=e-\frac{\uu^2}{2\rho}$). Total energy is $E=\int\diff\rr\,e(\rr)$, and $M=\int\diff\rr\, \rho$ is the total mass. Constant $T_0$ is the temperature an isolated system would have after relaxation to equilibrium and constant $\mu_0$ the equilibrium chemical potential. Local temperature is defined as the derivative of entropy with respect to the total energy density, i.e. $T^{-1}(\rr) =_\bydefinition S_{e(\rr)}$. Derivative of the MTL function with respect to momentum is then
\begin{equation}
  \uu^* = \left( \pd{\Psi^{(CH\rightarrow E)}}{\uu} \right)_{\rho,s} = S_\uu = -\frac{1}{T}  \frac{\uu}{\rho}.
\end{equation}
Assuming isothermal conditions, the conjugate momentum, $\uu^*$, is proportional to velocity $\vv = \uu/\rho$.

The dissipation potential is often in the form
\begin{equation}
  \Xi = \Xi \left( \Gamma(\uu^*) \right),
\end{equation}
$\Gamma$ being a linear operator. In our case 
\begin{equation}
  \label{eq:operator-gamma}
  \Gamma(\bullet) = -\frac{1}{2} \left(\nabla \bullet +(\nabla \bullet)^\top \right),
\end{equation}
which leads to shear rate $\Gamma(\uu^*)_{\uu^*=\Psi_\uu^{(CH\rightarrow E)}} = \frac{1}{T_0} \gradsym$. The constant prefactor $T_0$ is going to be ignored for the sake of simplicity. In a slightly overloaded notation, we can then write
\begin{equation}
  \Xi_{\uu^*} = \left\langle \Xi_{\XX},\pd{\XX}{\uu^*} \right\rangle 
  =\left\langle \Xi_{\XX}, \pd{\Gamma(\uu^*)}{\uu^*}\right\rangle
  =\transpose{\Gamma} \left(\Xi_{\XX}\right),
\end{equation}
where the adjoint operator $\transpose{\Gamma}$ is defined by
\begin{equation}
  \int \diff \rr\, \Gamma(\mathbf{a}) \cdot \mathbf{b} = \int \diff\rr\, \mathbf{a} \cdot \transpose{\Gamma} (\mathbf{b}), \qquad \forall \mbox{ symmetric tensor fields } \mathbf{a}, \mathbf{b}.
\end{equation}
This also specifies the scalar product $\langle\bullet,\bullet\rangle$ of two functions as the spatial integral of product of the functions ($L^2$ scalar product). In our case, we have $\transpose{\Gamma} (\bullet) = \divergence (\bullet)$. Eventually, the evolution equation for momentum density becomes
\begin{equation}\label{eq.evo.uu.Gamma}
  \pd{\uu}{t} = \left( \pd{\uu}{t} \right)_{rev} + \transpose{\Gamma} (\JJ),
\end{equation}
where $\JJ = \Xi_\XX$. Note that the shear stress $\dcstresssymb$ is a thermodynamic flux because it is conjugate to the force ($\XX$), which comes from differentiation of the MTL function (or entropy). Equation~\eqref{eq.evo.uu.Gamma}, which is the standard evolution equation for momentum density in fluid dynamics, is thus fully compatible with the GENERIC expression \eqref{eq.evo.uu.GENERIC}.

\subsection{Promoting conjugate stress to a new state variable}
\label{sec.promotion}

The goal is now to promote the flux to an independent state variable. But $\JJ$ can not be promoted itself because derivative of a functional with respect to $\JJ$ would then have to be on right hand side of evolution equation for $\uu$ instead of $\JJ$ itself. Rather, a conjugate flux $\JJconj$ is the new variable, that is coupled with momentum density. Since $\JJ$ is interpreted the shear stress, $\JJconj$ is interpreted as conjugate shear stress. As usual in non-equilibrium thermodynamics \cite{RedExt}, the dissipation can be moved onto a higher (more detailed) level of description, i.e. into the evolution equation for the extra state variable. Dissipative evolution of momentum density can be then recovered by letting the extra variable relax. The reversible coupling between the original state variables and the extra state variable is required to be antisymmetric in order to fulfill the Onsager--Casimir reciprocal relations \cite{Casimir1945,PRE2014,HCO} and thus we write (still following \cite{MG-CR}) evolution equations for extended hydrodynamics (EH) as follows
\begin{subequations}
  \label{eq.EH.evo}
  \begin{eqnarray}
    \label{eq.u.evo.EH}
    \pd{\uu}{t} &=& \left( \pd{\uu}{t} \right)_{rev} + \transpose{\Gamma} (\JJconjconj),\\
    \label{eq.JJc.evo}
    \pd{\JJconj}{t} &=& - \Gamma(\uu^*) + \Xi^*_{\JJconjconj}.
  \end{eqnarray}
\end{subequations}
To close the evolution equations, the conjugates are evaluated as the corresponding derivatives of a new extended MTL function $\Psi^{(EH \to E)}$, $\uu^* = \Psi^{(EH \rightarrow E)}_{\uu}$ and $\JJconjconj = \JJ = \Psi^{(EH \rightarrow E)}_{\JJconj}$. Note that $\JJconjconj$ was identified with $\JJ$ because the generalized Legendre transformation is idempotent (second application of the transformation restores the original functional), i.e. $\JJconjconj = \JJ$. The MTL function $\Psi^{(EH \rightarrow E)}$, which drives evolution of the extended hydrodynamics towards equilibrium, has yet to be specified in order to define the conjugate stress tensor completely. Evolution equations \eqref{eq.EH.evo} represent evolution on the level of extended hydrodynamics.

Parity with respect to the time-reversal \cite{PRE2014} of $\JJ$ is now even, which is in contrast to the odd parity of $\JJ$ before promoting $\JJconj$ to a state variable. The physical reason is that we have now a more detailed description of the shear stress, which can be regarded for example as the stress formulated by kinetic theory (App.~\ref{sec.KT}). Similarly, $\JJconj$ and $\JJconjconj$ are even with respect to time reversal, since MTL functions and dissipation potentials, with respect to which the fluxes are conjugate, are always even. Therefore, evolution equation \eqref{eq.u.evo.EH} is now reversible (in contrast to Eq.~\eqref{eq.u.evo}) while equation \eqref{eq.JJc.evo} contains a reversible term (first) and an irreversible term (second). By moving to a more detailed level of description, parity of the quantity promoted to an extra state variable may change because the quantity becomes independent and is no longer enslaved by the constitutive relation valid before the promotion. As a consequence, irreversible terms may become reversible.

\subsection{Specification of the MTL function}
The conjugate stress will be now specified more precisely. The relation between $\JJ$ and $\JJconj$ is a Legendre transformation,
\begin{equation}
  \label{eq.LT.JJ.JJc}
  \JJ = \Psi^{(EH\rightarrow E)}_{\JJconj}, \quad \text{ or } \quad
  \pd{}{\JJconj} \left( - \Psi^{(EH\rightarrow E)} + \int \diff \rr\, \JJ : \JJconj \right) = \tensorq{0},
\end{equation}
which is carried out with respect to the MTL function 
\begin{equation}\label{eq.Psi.EH.E}
  \Psi^{(EH \rightarrow E)} =_\bydefinition S^{(EH\rightarrow E)} - \frac{1}{T_0}E^{(EH\rightarrow E)} + \frac{\mu_0}{T_0}M^{(EH\rightarrow E)}.
\end{equation} 
This MTL function drives the extended hydrodynamics towards equilibrium.  Note that all quantities with $(EH\rightarrow E)$ superscript depend on the hydrodynamic fields $\xx^{(CH)} =_\bydefinition (\rho,\uu,e)$ and the conjugate stress tensor $\JJconj$. Quantities $E^{(EH\rightarrow E)}$, $S^{(EH\rightarrow E)}$ and $M^{(EH\rightarrow E)}$, which have so far been left undetermined, are the building blocks of the corresponding MTL function, and they can reflect the particular nature of the physical system. Alternatively, the extra evolution equation can be interpreted as a numerical scheme leading to constitutive relation \eqref{eq:le-roux-rajagopal}.

Let us consider at least one possible choice of $\Psi^{(EH\rightarrow E)}$. Within the kinetic theory (App.~\ref{sec.KT}), kinetic energy of particles is equal to the sum of kinetic energy of the overall motion $\uu^2/(2\rho)$ and pressure, and the deviatoric part of stress tensor thus does not contribute to energy (when neglecting long-range pair interactions). On the other hand, additional knowledge expressed as the extra state variable reduces entropy, and the entropy on the $(EH)$ level is thus lower than the entropy on the level of classical hydrodynamics $(CH)$. Similarly as in Extended Irreversible Thermodynamics \cite{Jou-EIT}, the first approximation of the $(EH)$ entropy is
\begin{equation}
  \label{eq.EH.S}
  S^{(EH)} = S^{(CH)} -\int \diff\rr\, \frac{1}{2}\beta \absnorm{\JJconj}^2 = \int\diff\rr\, s\left(\rho,e-\frac{\uu^2}{2\rho}\right) - \frac{1}{2}\beta \absnorm{\JJconj}^2.
\end{equation}
Coefficient $\beta$ is positive so that the entropy is concave. Energy and mass on the $(EH)$ level are the same as on the $(CH)$ level, thus the MTL function driving evolution from $(EH)$ to equilibrium is
\begin{equation}
  \Psi^{(EH\rightarrow E)} = \Psi^{(CH\rightarrow E)} -\int\diff\rr\, \frac{1}{2} \beta\absnorm{\JJconj}^2,
\end{equation}
from which it follows that
\begin{equation}\label{eq.T.Tc}
  \JJ = \Psi^{(EH\rightarrow E)}_{\JJconj} = - \beta \JJconj.
\end{equation}
In summary, for the particular choice of the entropy \eqref{eq.EH.S}, the conjugate stress tensor $\JJconj$ is proportional to the standard shear stress tensor $\dcstresssymb$   and the evolution equations~\eqref{eq.EH.evo} for the constitutive relation~\eqref{eq:le-roux-rajagopal} and the operator $\Gamma$ defined by~\eqref{eq:operator-gamma} read
\begin{subequations}
  \begin{align}
    \pd{\vecu}{t} &= \left( \pd{\vecu}{t} \right)_{rev} + \divergence \dcstresssymb, \\
    - \frac{1}{\beta} \pd{\dcstresssymb}{t} &= - \gradsym + \left[ \alpha (1 + \beta \absnorm{\dcstresssymb}^2)^n + \gamma \right] \dcstresssymb. \label{eq:26b}
  \end{align}
\end{subequations}
The second equation can be apparently interpreted as the time derivative regularization of the original constitutive relation.

\subsection{Duality between $\JJ$ and $\JJconj$}
\label{sec.EH.duality}

The inverse Legendre transformation from $\JJconj$ to $\JJ$ follows. The conjugate MTL function $\Psi^{*,(EH\rightarrow E)}$ can be constructed as
\begin{equation}
  \Psi^{*,(EH\rightarrow E)} (\uu^*,\JJ) = - \Psi^{(EH\rightarrow E)} \left( \uu, \JJconj(\JJ) \right) + \int \diff\rr\, \JJconj(\JJ) : \JJ
\end{equation}
where $\JJconj(\JJ)$ is the solution to \eqref{eq.LT.JJ.JJc}, i.e. Eq. \eqref{eq.T.Tc} for the particular choices of entropy and energy. Conversely, the Legendre transformation
\begin{equation}
  \label{eq.LT.JJc.JJ}
  \JJconj = \Psi^{*,(EH\rightarrow E)}_{\JJ}, \quad \text{ or } \quad
  \frac{\partial}{\partial \JJ}\left(-\Psi^{*,(EH\rightarrow E)} + \int \diff \rr\, \JJ : \JJconj \right) = \tensorq{0}
\end{equation}
leads back to the original MTL function,
\begin{equation}
  \Psi^{(EH\rightarrow E)} (\uu, \JJconj) = -\Psi^{*,(EH\rightarrow E)} \left( \uu^*, \JJ(\JJconj) \right) + \int \diff \rr\, \JJ(\JJconj):\JJconj,
\end{equation}
where $\JJ(\JJconj)$ are the solutions to Eq.~\eqref{eq.LT.JJc.JJ}. The Legendre duality between $\JJ$ and $\JJconj$ has thus been established.

\subsection{Fast reducing evolution}
\label{sec.fast}

Now we are coming to the highlight of the thermodynamic extension. Let us now identify the evolution of $\JJconj$ that leads back to classical hydrodynamics, i.e. the reducing (fast) evolution leading to the reduced (hydrodynamic) evolution of variables $\xx$. Equation \eqref{eq.JJc.evo} can be rewritten as
\begin{equation}
  \label{eq.reducing}
  \pd{\JJconj}{t} = - \pd{}{\JJ} \left( \left\langle \Gamma(\uu^*), \JJ \right\rangle - \Xi^* \right) \Big|_{\left(\uu^* = \Psi^{(EH \rightarrow E)}_{\uu}, \JJ= \Psi^{(EH \rightarrow E)}_{\JJconj}\right)} = - \Psi^{(EH\rightarrow CH)}_{\JJ},
\end{equation}
where
\begin{equation}
  \label{eq.Psi.EH.CH}
  \Psi^{(EH\rightarrow CH)}(\XX,\JJ) =_\bydefinition -\Xi^*(\JJ) + \int \diff \rr\, \XX : \JJ,
\end{equation}
with $\XX = \Gamma(\uu^*)$, is the MTL function driving evolution from extended hydrodynamics to classical hydrodynamics. This MTL function will be referred to as the \textbf{reducing MTL function}. Equation \eqref{eq.reducing} then means that conjugate shear stress $\JJconj$ evolves in the direction of steepest ascent of $\Psi^{(EH\rightarrow CH)}(\XX,\JJ)$. MTL function \eqref{eq.Psi.EH.CH} generates Legendre transformation from $\Xi^*$ to $\Xi$ as 
\begin{equation}
  \Psi^{(EH\rightarrow CH)}_\JJ = \tensorq{0},
\end{equation}
which is equivalent to Eq.~\eqref{eq.Xic.Xi}. The constitutive relation is thus recovered after the extra variable ($\JJconj$) has relaxed to a state enslaved by the slow variable ($\uu$). 

The reducing evolution \eqref{eq.reducing} can be alternatively rewritten as
\begin{equation}\label{eq.reducing.J}
  \pd{\JJ}{t} = - \left\langle \frac{\partial^2 \Psi^{(EH\rightarrow E)}}{\partial \JJconj \partial \JJconj}, \Psi^{(EH\rightarrow CH)}_{\JJ} \right\rangle,
\end{equation}
where Eq.~\eqref{eq.LT.JJ.JJc} was used, and compatibility of the reducing evolution \eqref{eq.reducing} with the second law of thermodynamics is guaranteed by 
\begin{equation}
  \dot{\Psi}^{(EH\rightarrow CH)} = -\int \diff \rr\, \Psi^{(EH\rightarrow CH)}_{\JJ} \Psi^{(EH\rightarrow E)}_{\JJconj \JJconj} \Psi^{(EH\rightarrow CH)}_{\JJ} \geq 0,
\end{equation}
for $\Psi^{(EH\rightarrow E)}$ concave. Note that only derivatives with respect to the fast variable are taken into account because only evolution of the fast variable is considered in the reducing evolution. The MTL function $\Psi^{(EH\rightarrow E)}$ has to be thus chosen concave, and the reducing evolution is compatible with the second law of thermodynamics.

On the other hand, dissipation potentials $\Xi$ and $\Xi^*$ were not convex and thus 
\begin{equation}
  \dot{\Psi}^{(CH\rightarrow E)} = \int \diff \rr\, \Psi^{(CH\rightarrow E)}_{\uu} \left. \Xi^*_{\uu^*} \right|_{\uu^* = \Psi^{(CH\rightarrow E)}}
\end{equation}
does not need to have definite sign and evolution on the hydrodynamic level can have apparent negative entropy production. Indeed, if shear rate and shear stress were spatially homogeneous with only for example $xy$ component nonzero, the slope of $\Xi^*$ could be negative for positive $\tensorc{D}_{xy}$. The way out of such a problem is the lift to extended hydrodynamics, where entropy production keeps its positivity. In other words, apparent negative entropy production would indicate evolution of instability (phase transition), and can be regarded as an invitation to a higher (more detailed) level of description.

After having demonstrated how the constitutive relation can be interpreted as a consequence of a more detailed evolution, study of behavior of the more detailed evolution gives information about CR-stability of the evolution with respect to the constitutive relation itself. The constitutive relation is stable if any perturbation decreases in time. It is metastable if only small perturbations decrease in time, and it is unstable if no perturbation decreases in time. The relation between CR-stability and the usual mathematical concept of stability is discussed in \ref{sec.stab-CRstab}.

\subsection{Alternative forms of the extension}

Evolution equations \eqref{eq.EH.evo} are relatively simple and useful for qualitative analysis of CR-stability carried out in Sec. \ref{sec.stability}, but they are not objective due to the presence of partial time derivative of the conjugate stress instead of an objective derivative. Moreover, Jacobi identity is not fulfilled for the system of equations (checked by program \cite{Kroeger2010}). Hence, the evolution equations have to be regarded rather as a toy model useful for explaining construction of the extension and the CR-stability implications.

The extension can be made more physically sound by choosing a better operator $\Gamma$. Such a choice, however, must be accompanied by an another choice of the quantity to be promoted to an extra state variable. For example, the choice
\begin{equation}
  \label{eq.Gamma.c}
  \divergence \cstress= \Gamma^T(\cc^*) = T_0 c_{jk}\partial_i c^*_{jk} -T_0 \partial_k\left( c_{kj} (c^*_{ij}+c^*_{ji})\right)
\end{equation}
with $\cc$ being the new variable (conformation tensor) leads to the adjoint operator 
\begin{equation}
  \Gamma(\uu^*) = -T_0\partial_k(c_{ij} u^*_k) + T_0 c_{kj}\partial_k u^*_i + c_{ki}\partial_k u^*_j.
\end{equation}
It should be stressed that this extension is considered in the energetic representation, where entropy density is among the state variables instead of the total energy density. Let us refer to this extension as to the conformation tensor (CT) level. 

The evolution equation for the extra variable is then
\begin{equation}
  \label{eq.evo.c}
  \frac{\partial c_{ij}}{\partial t} = -\Gamma(\uu^*) + \frac{\partial \Xi^*}{\partial c^*_{ij}}
  =\partial_k(c_{ij} T_0 u^*_k) - c_{kj}T_0 \partial_k u^*_i - c_{ki}T_0 \partial_k u^*_j + \frac{\partial \Xi^*}{\partial c^*_{ij}},
\end{equation}
which is analogical to Eq. \eqref{eq.EH.evo}. The reversible part is the upper-convected derivative of $\cc$ (with an extra compressibility term, which is sometimes referred as to the Truesdell rate). Indeed, in the energetic representation total entropy depends only on the entropy density and derivative of MTL function \eqref{eq.Psi.EH.E} becomes
\begin{equation}
  \Psi^{(CT\rightarrow E)}_{\uu} = -\frac{1}{T_0} E_\uu = -\frac{1}{T_0}\frac{\uu}{\rho},
\end{equation}
and Eq. \eqref{eq.evo.c} then becomes
\begin{equation}
  \frac{\partial c_{ij}}{\partial t} 
  + \partial_k(c_{ij} v_k) - c_{kj} \partial_k v_i - c_{ki} \partial_k v_j =  \frac{\partial \Xi^*}{\partial c^*_{ij}},
\end{equation}
where the left hand side is the upper-convective derivative of $\cc$. Note that velocity is given by $\uu/\rho$. Such an extended evolution also fulfills the Jacobi identity, is thus Hamiltonian, see e.g. \cite{Gn1}, and is also objective. Objectivity seems to be in tight relation with validity of Jacobi identity. Evolution equation \eqref{eq.evo.c} is the analogy of Eq.~\eqref{eq.EH.evo} and this whole paragraph is the analogy of Sec.~\ref{sec.promotion}.

Conjugate dissipation potential $\Xi^*$ is defined through Legendre transformation
\begin{subequations}
  \begin{equation}
    \pd{}{\uu^*} \left(-\Xi(\uu^*) + \langle \uu^*, \Gamma^T(\cc^*)\rangle\right) = \tensorq{0} \quad\Rightarrow\quad \uu^* = \uu^*(\cc^*),
  \end{equation}
  and 
  \begin{equation}
    \Xi^*(\cc^*) = - \Xi \left( \uu^*(\cc^*) \right) + \langle \Gamma(\uu^*(\cc^*)), \cc^*\rangle,
  \end{equation}
\end{subequations}
which is analogical to the transformations carried out in Sec. \ref{sec.conjugate}.

The relation between $\cc$ and $\cc^*$ can be obtained as in Sec.~\ref{sec.EH.duality}, i.e. through Legendre transformation with respect to a MTL function $\Psi^{(CT\rightarrow E)}$. The MTL function is constructed from entropy, energy and total mass (as in Eq.~\eqref{eq.Psi.EH.E}), see e.g. \cite{Sarti-Marrucci,Gn1}. A particular choice of entropy could be
\begin{equation}
  \label{eq.CT.S}
  S (\rho, \uu, e, \cc)^{(CT)} = \int \diff\rr\, s \left( \rho,e-\frac{\uu^2}{2\rho} - H \Tr(\cc)\right) + \frac{1}{2}k_B\frac{\rho}{m} \ln \det \frac{\cc}{Q} - \frac{1}{2}\beta \partial_i c_{jk} \partial_i c_{jk},
\end{equation}
where $H$, $Q$ and $\beta$ are some material-dependent parameters and the last term expresses weakly non-local effects as in \cite{Vitek-stress-diffusion}. The weakly non-local effects could play a role when the conformation tensor should exhibit steep spatial variations, as is usual in shear banding. Entropy \eqref{eq.CT.S} is just one possible choice made here for being explicit. However, the concrete form of entropy depends on the class of materials considered.

Finally, Eq.~\eqref{eq.evo.c} can be rewritten as
\begin{equation}
  \pd{\cc}{t} = - \pd{}{\cc^*} \Psi^{(CT \rightarrow CH)},
\end{equation}
where the reducing MTL function is 
\begin{equation}
  \Psi^{(CT\rightarrow CH)} = -\Xi^*(\cc^*) + \langle \Gamma(\uu^*), \cc^*\rangle.
\end{equation}
This is completely analogical to Sec.~\ref{sec.fast}, and analysis of the second law of thermodynamics could proceed as in that section.

This alternative extension brings this paper closer to the common modeling of shear banding, where viscoelastic models are usually employed (see the discussion in Sec.~\ref{sec.CR}). Similarly, by choosing a different operator $\Gamma$, one can derive evolution equations for Reynolds stress or weakly non-local vorticity, that can be found in \cite{Miroslav-turbulence,PhysicaD-2015}. In the following section, however, we stick to the extension where conjugate stress tensor is the extra state variable for simplicity.

\section{Interpretation of CR-stability}
\subsection{Thermodynamic interpretation}\label{sec.stability}

\begin{figure}[!htbp]
  \centering
  \includegraphics[width=0.6\textwidth]{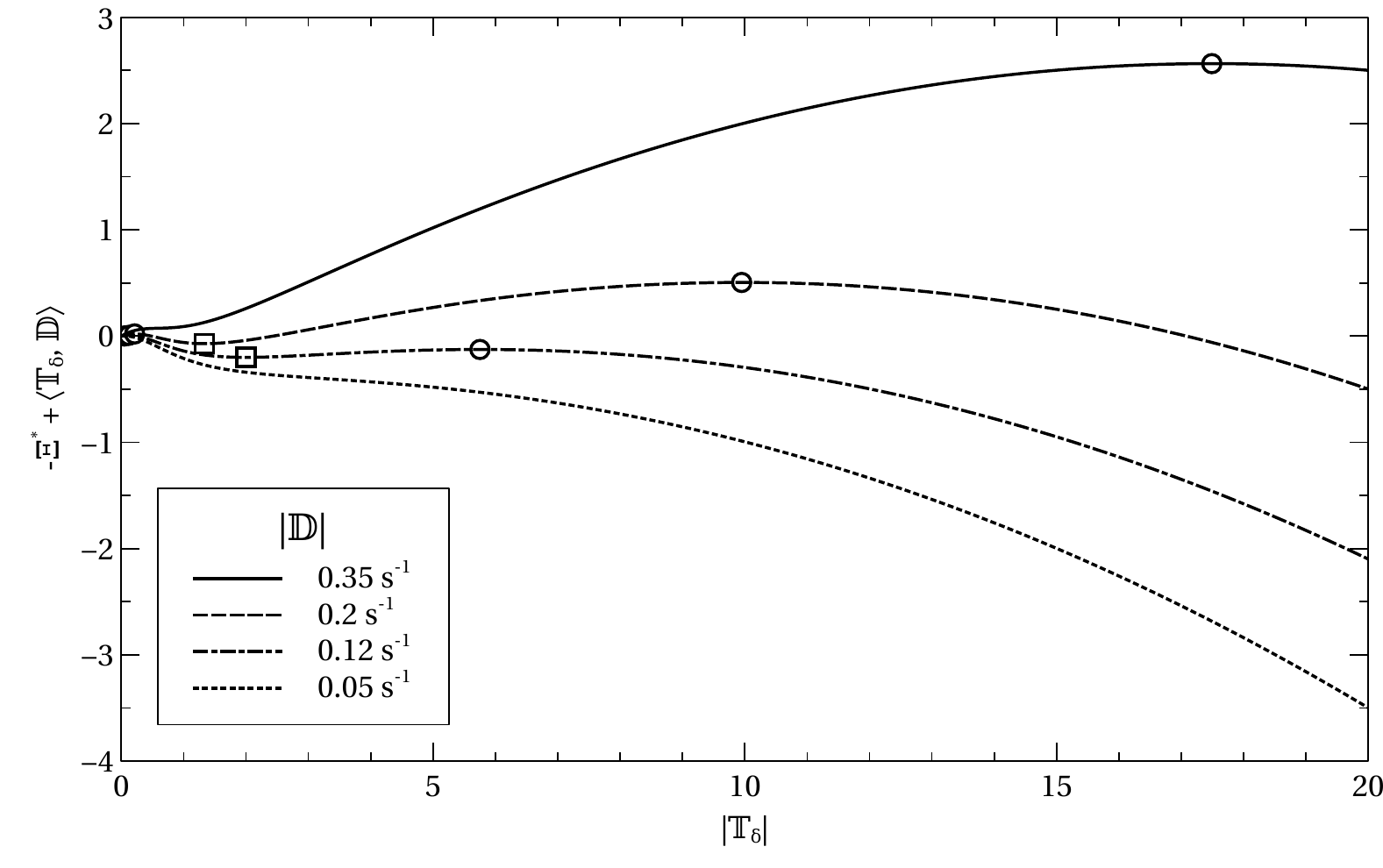}
  \caption{Reducing MTL function $\Psi^{(EH\rightarrow CH)}$, Eq.~\eqref{eq.Psi.EH.CH}. For low shear rates $\absnorm{\gradsym}$ (thermodynamic force $\XX$) the function has only one local maximum, which is located near the origin ($\absnorm{\dcstresssymb} = \unit[0]{Pa}$). As the shear rate increases, a second local maximum develops in regions with higher shear stress $\dcstresssymb$. As $\absnorm{\gradsym}$ further increases, the second maximum becomes higher than the first maximum. Finally, for even higher shear rates, the first maximum disappears. Local maxima are indicated by circles ($\circ$) while local minima by squares ($\square$).}
  \label{fig:slope-transform}
\end{figure}

The reducing MTL function $\Psi^{(EH\rightarrow CH)}$ is plotted in Fig.~\ref{fig:slope-transform} for several values of shear rates $\XX$.

The curves plotted in the figure correspond to the following different regimes of the reducing evolution, indicated also in Fig.~\ref{fig:conjugate-dissipation-potential}. 
\begin{enumerate}[I]
\item For low shear rates $\absnorm{\gradsym}$ (Regime I when comparing to Fig. \ref{fig:conjugate-dissipation-potential}) there is only one local maximum, which is near the origin ($\absnorm{\dcstresssymb} = \unit[0]{Pa}$), as the curve for $\absnorm{\gradsym} = \unit[0.05]{s^{-1}}$ indicates. Bearing in mind that $\Psi^{(EH\rightarrow E)}$ has to be chosen concave, Eq.~\eqref{eq.reducing.J} tells that shear stress evolves in the direction of derivative of the reducing MTL function. In other words, if the derivative is positive, $\dcstresssymb$ increases, and if it is negative, $\dcstresssymb$ decreases. Thus, if $\dcstresssymb$ is initially to the left of the local maximum, where the derivative is positive, it tends to the local maximum. If it is to the right of the maximum, where the derivative is negative, it again tends to the maximum. $\dcstresssymb$ is thus attracted to the local (and global) maximum. The evolution towards this maximum is thus CR-stable.
\item For slightly higher shear rates (Regime II), another local maximum appears in higher shear stresses (curve for $\absnorm{\gradsym} = \unit[0.12]{s^{-1}}$). Similarly as in Regime I (low shear rates), this new local maximum attracts $\dcstresssymb$ from its neighborhood. However, since the first maximum is still higher, settling of $\dcstresssymb$ in the second maximum can be seen as a CR-metastable state. After a sufficient fluctuation, $\dcstresssymb$ will end up in the first (higher) maximum. Moreover, a local minimum appears between the local maxima. The local minimum is an CR-unstable state, since $\dcstresssymb$ is attracted to the local maxima regardless how close to the local minimum it is. That can be seen from signs of derivative of the reducing MTL function to the left and to the right of the local minimum.
\item For even higher shear rates (Regime III and curve $\absnorm{\gradsym} = \unit[0.2]{s^{-1}}$), the second maximum becomes higher than the first one, and the first maximum thus becomes CR-metastable while the second becomes CR-stable. The local minimum is still located between the maxima, and although it is a stationary point ($\dot{\cstress}_\delta = \tensorq{0}$ in the local minimum), such a state is CR-unstable, since any arbitrarily small fluctuation in the stress would make the stress go towards one of the maxima. Note that the value of the shear rate between Regimes II and III, when the two local maxima have the same height, corresponds to the shear rate selected by the Maxwell lever rule (see Sec. \ref{sec.CR}).
\item Finally, (Regime IV) the first local maximum disappears (as well as the local minimum) for even higher shear rates (curve $\absnorm{\gradsym} = \unit[0.35]{s^{-1}}$), and the second maximum becomes the only CR-stable state.
\end{enumerate}

The described dynamics of the stress tensor can be seen also in Fig.~\ref{fig:conjugate-dissipation-potential}. Regime I (low shear rates) corresponds to the part of curve $A-B$ left of point $D$, where only one $\dcstresssymb = \Xi_\gradsym$ is possible. Regime II (one CR-stable state, one CR-unstable state and one CR-metastable state) is indicated in that figure as well. The upper solid curve is the CR-stable state, the lower solid curve is the CR-metastable state and the dashed curve is the CR-unstable state. Regime III (one CR-metastable state, one CR-unstable state and one CR-stable state) is described analogically. Finally, Regime IV has only one state, which is CR-stable. The reducing evolution \eqref{eq.reducing.J} is thus able to distinguish between CR-stable, CR-metastable and CR-instable states in Fig.~\ref{fig:conjugate-dissipation-potential}, and the analysis is similar to the treatment of phase transitions in equilibrium thermodynamics \cite{Landau5,Callen}. We are thus coming to a kind of \textbf{dissipative phase transitions}.

Let us now return to the experimental results in Fig.~\ref{fig:conjugate-dissipation-potential} and to the second question raised in the introduction. When slowly raising the shear rate, we start in Regime I, where shear stress is determined uniquely. After passing to Regime II, an another possible state appears, but shear stress evolves continuously due to carefulness of the experimentalists (smallness of fluctuations). Passing also through Regime III, point $C$ (boundary between Regimes III and IV) is reached, where the CR-metastable state becomes CR-unstable and shear stress jumps to the CR-stable state. That is why \textbf{shear stress behaves discontinuously}.

Moreover, it could be expected that when going back from high shear rates to the small ones, shear stress would fall as low as to point $D$ in Fig.~\ref{fig.exp}, from which it would jump to the CR-stable state on branch $A$. One should thus observe \textbf{hysteresis} indicated in that figure. We are not aware of such a backward experiment having been carried out, but modeling of hysteresis by means of introducing a new evolution equation is usual in literature \cite{kubin.lp.poirier.jp:relaxation}.

On the other hand, when slowly varying $\dcstresssymb$, there is always a unique value of $\gradsym$, determined by slope of dissipation potential $\Xi^*$ in Fig.~\ref{fig:dissipation-potential}, and the slope varies continuously. That is why \textbf{no jump in shear rate} is observed in the experiment.

  \subsection{Stability of the regularized constitutive relation}
  \label{sec.stab-CRstab}
  
  The same dynamical behavior can be concluded directly from the stability analysis of the regularized constitutive relation \eqref{eq:26b}. Its one-dimensional counterpart with a general conjugate dissipation potential $\Xi^*$ reads
  \begin{equation}
    \label{eq:45}
    \frac{1}{\beta} \pd{\absnorm{\dcstresssymb}}{t} = \absnorm{\gradsym} - \Xi^*_{\absnorm{\dcstresssymb}} \left(\absnorm{\dcstresssymb} \right).
  \end{equation}

\begin{figure}[!htbp]
    \centering
    \includegraphics[width=0.8\textwidth]{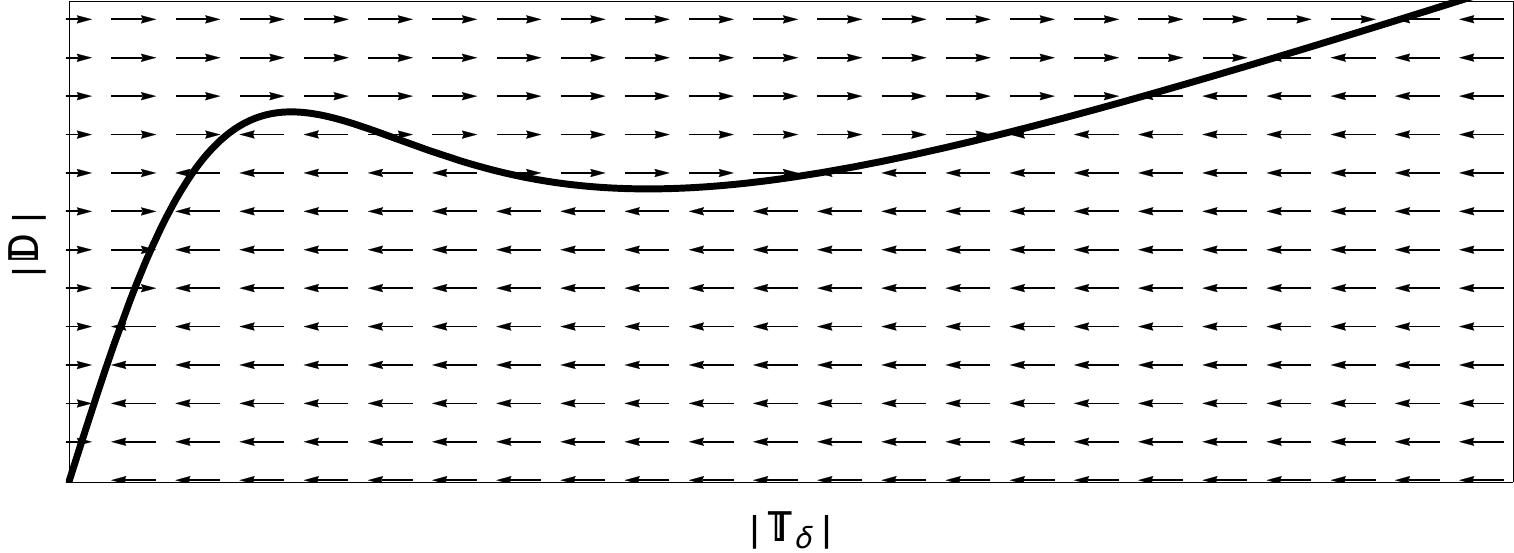}
    \caption{Constitutive relation generated by the dual dissipation potential \eqref{eq.theta} with parameters $\alpha = 1$, $\beta = 1$, $\gamma = 0.12$ and $n = -2$, and the corresponding phase field. Vectors are rescaled to have uniform size.}
    \label{fig:phase-field}
  \end{figure}

In the shear-rate-controlled experiment we assume that $\absnorm{\gradsym} = \absnorm{\gradsym} (t)$ is given. Equation \eqref{eq:45} is then a non-autonomous ordinary differential equation with stationary points $\absnorm{\gradsym} - \Xi^*_{\absnorm{\dcstresssymb}} \left(\absnorm{\dcstresssymb} \right)$. It can equivalently written as an extended autonomous system
  \begin{subequations}
    \begin{align}
      \pd{\absnorm{\dcstresssymb}}{t} &= \beta \left[ \absnorm{\gradsym} (\tau) - \Xi^*_{\absnorm{\dcstresssymb}} \left(\absnorm{\dcstresssymb} \right) \right], \\
      \pd{\tau}{t} &= 1.
    \end{align}
  \end{subequations}
  From this system, we can deduce the phase field in the $\absnorm{\gradsym}$--$\absnorm{\dcstresssymb}$ plane. In Figure~\ref{fig:phase-field}, we show the phase field for the particular choice of the dual dissipation potential \eqref{eq.theta}.
  We can easily see that the solutions are quickly (depending on the parameter $\beta$) attracted towards the increasing parts of the constitutive relation and repelled from the decreasing part. In other words, the decreasing part is unstable in the usual mathematical sense while the increasing parts are asymptotically stable. This leads to the jumps (and possibly hysteresis) in the values of the stress $\absnorm{\dcstresssymb}$ when changing $\absnorm{\gradsym}$.

 On the other hand, if we prescribe $\absnorm{\dcstresssymb} = \absnorm{\dcstresssymb} (t)$ (shear stress controlled experiment) and we further assume that its time derivative is small enough, \eqref{eq:45} reduces back to the original constitutive relation. In this case, there are no jumps.

At least in this case the CR-stability analysis leads to the same conclusion as the usual stability analysis for ordinary differential equations provided the evolution of the extra state variable (here conjugate stress) is assumed much faster than evolution of the remaining variables.

\subsection{Linearized hydrodynamic stability}
\label{sec:linearized-stability}

The CR-instability of the decreasing part of the constitutive relation can be also verified by means of the linearized hydrodynamic stability. Perturbing a known simple shear base flow, the aim is to study the temporal evolution of the disturbance. Linearizing the governing equations for the disturbance yields a generalized eigenvalue problem. Since constitutive relation \eqref{eq:le-roux-rajagopal} does not have to be invertible, the resulting problem might be in the form of a system of equations, unlike its classical counterpart---the single Orr--Sommerfeld equation, see Orr \cite{orr.w:stability*1,orr.w:stability} and Sommerfeld \cite{sommerfeld.a:beitrag}. Using a variant of a pseudospectral collocation method, e.g. Canuto and Hussaini \cite{canuto.c.hussaini.my.ea:spectral*1,canuto.c.hussaini.my.ea:spectral}, we are able to solve the problem and numerically confirm that once the flow is governed by a non-monotone constitutive relation, the perturbation grows in time and the flow is unstable. The derivation of the problem and description of the numerical method is rather lengthy and it will be subject of a future publication.

The hydrodynamic stability is usually studied for a Poiseuille or Couette flow of a fluid in some archetypal geometry like a two-dimensional channel or a pipe with prescribed boundary conditions. In the classical setting, this yields a single differential eigenvalue equation, while in our case, it leads to a system of equations. In either case, it has to be solved numerically and the spectral collocation method has proven effective, see \cite{Orszag}. This is different from, for example, Ván \cite{Van-stability} who could use the Routh--Hurwitz criteria, since he studied the stability of the equilibrium and he considered the perturbation in the form of an unbounded plane wave, resulting in a third-order polynomial. We cannot make use of the mentioned criteria as we want to study stability of non-equilibrium flows with imposed perturbations also satisfying the boundary conditions, and consequently we do not obtain just a polynomial. Anyway, determining the stability of the equilibrium can be also done from the kinetic energy of the disturbance.

\section{Braun--Le Chatelier principle}

Braun--Le Chatelier principle in equilibrium thermodynamics is the statement that a perturbation of a thermodynamic force acting on a system triggers such internal processes inside the system that the effect of the perturbation is reduced after the processes have relaxed \cite{LeChatelier}. The principle can be derived from convexity of the energy functional \cite{Landau5}, and it is equivalent with stability of the system. Analogical derivation starting from convexity of a dissipation potential leads to an analogical result, Braun--Le Chatelier principle for dissipative thermodynamics \cite{Braun-LeChat}, 
\begin{equation}
  \left( \pd{\absnorm{\dcstresssymb}}{\absnorm{\gradsym}} \right)_{J_{int}=0} < 
  \left( \pd{\absnorm{\dcstresssymb}}{\absnorm{\gradsym}} \right)_{X_{int}},
\end{equation}
where $J_{int}$ is a rate of change of an internal variable and $X_{int}$ is the thermodynamic force corresponding on the internal variable. In other words, apparent viscosity (the derivative of stress with respect to strain) is lower after the internal variable has relaxed ($J_{int}=0$) than just after the perturbation (change of shear rate). 

Therefore, Braun--Le Chatelier principle, which is a consequence of relaxation of an internal variable and convexity of a dissipation potential, leads to shear thinning. This is in agreement with that shear thinning is usually observed in the gradient banding experiments, which are well described by a viscoelastic model. The internal variable dynamics of which leads to the viscoelastic behavior relaxes and causes shear thinning in agreement with the Braun--Le Chatelier principle. 

On the other hand, shear thickening is more difficult to explain, which supports the use of non-linear or even non-convex dissipation potentials.

\section{Non-smooth dissipation potentials}

This section stands a little apart from the rest of the paper, as it addresses yield stress behavior instead of S-curves. It is included because the authors believe that discussing non-smoothness of dissipation potentials is a natural continuation of the discussion of non-convexity.

\begin{figure}[!htbp]
  \centering
  \begin{minipage}[t]{0.48\textwidth}
  \includegraphics[width=\textwidth]{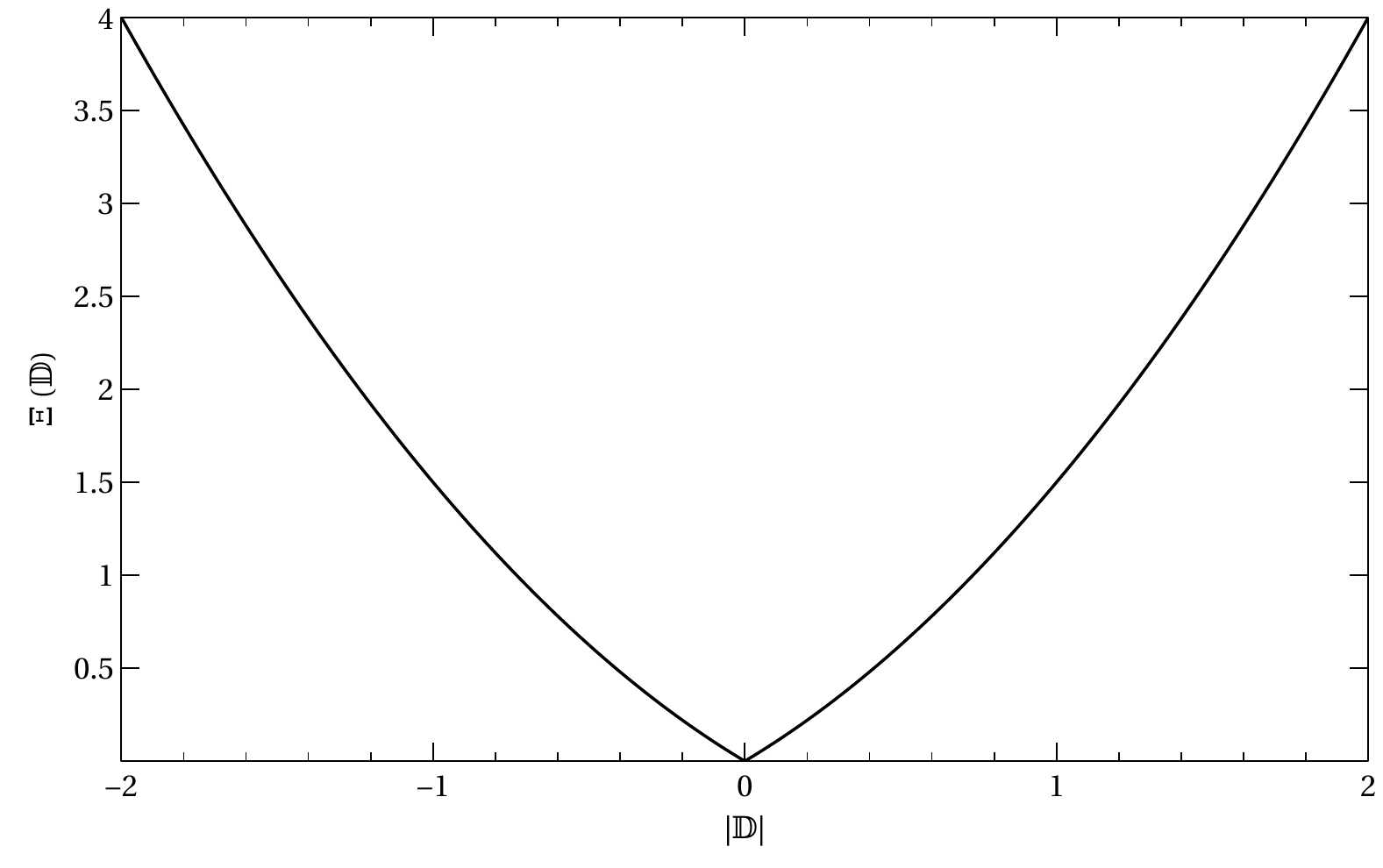}
  \caption{Dissipation potential \eqref{eq.Bingham.Xi} exhibiting yield stress. The potential is not differentiable at $\absnorm{\gradsym} = 0$. Numerical values of $\alpha$ and $\beta$ are taken as 1.}
  \label{fig.Bingham.Xi}
  \end{minipage}
  \hfill
  \begin{minipage}[t]{0.48\textwidth}
  \includegraphics[width=\textwidth]{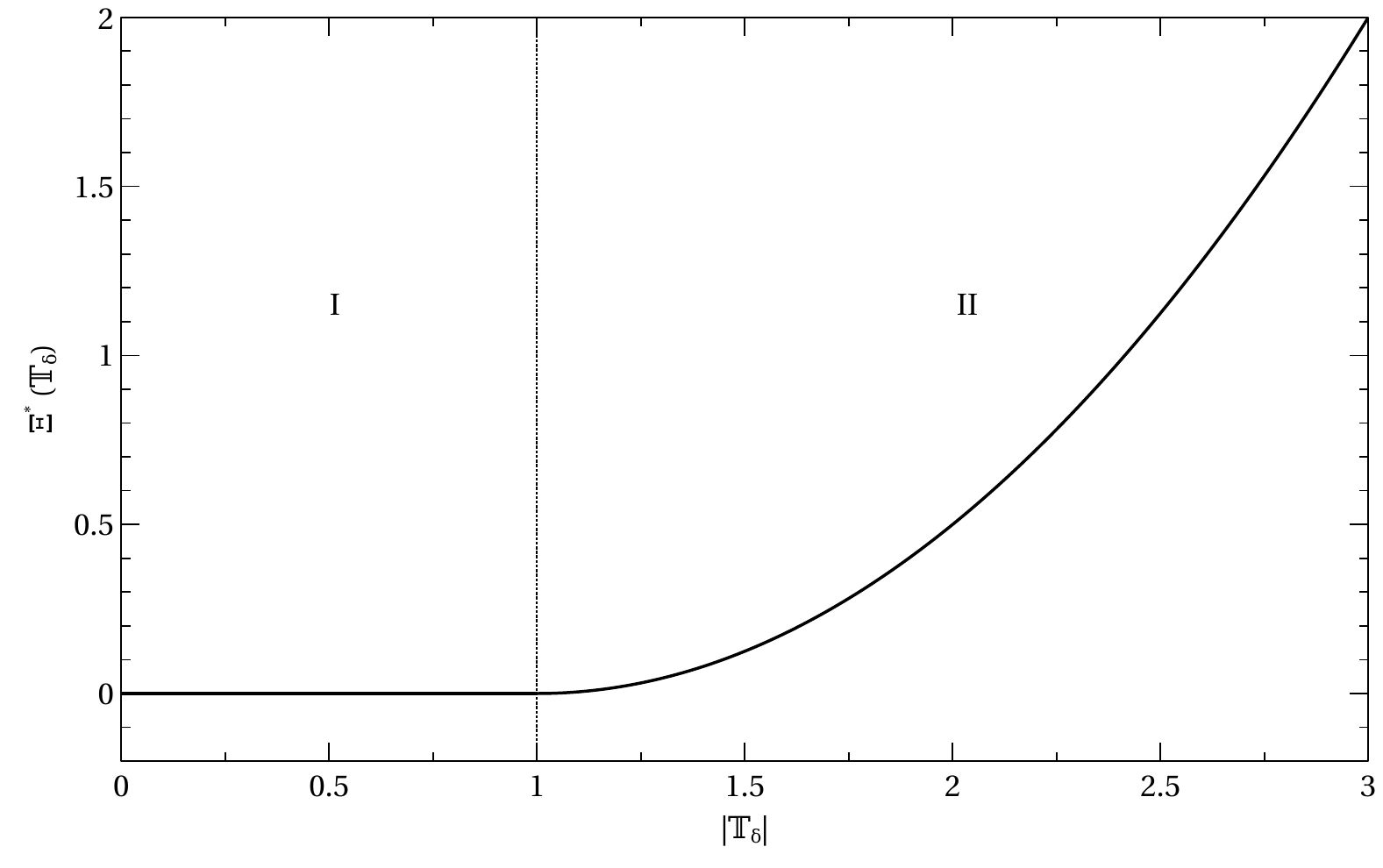}
  \caption{Conjugate dissipation potential $\Xi^*$ obtained by Legendre transformation using subdifferentials from~\eqref{eq.Bingham.Xi}. Note that it is smooth everywhere in contrast to $\Xi$. Conjugate potential is flat near the origin and grows for higher shear stresses.}
  \label{fig.Bingham.Xic}
  \end{minipage}
\end{figure}

Inspired by~\cite{Rajagopal2004}, dissipation potential exhibiting yield stress (no shear rate for sufficiently low shear stress) is for example
\begin{equation}
  \label{eq.Bingham.Xi}
  \Xi(\gradsym) = \int \diff\rr\, \alpha \absnorm{\gradsym} + \frac{1}{2}\beta \absnorm{\gradsym}^2,
\end{equation}
where $\alpha$ and $\beta$ are positive constants, see Fig.~\ref{fig.Bingham.Xi}. The yield stress (given by the coefficient $\alpha$) is caused by the fact that for low shear stresses (low slopes of the dissipation potential) there is no corresponding $\gradsym$. More exactly, the only shear rate corresponding to low shear stresses is zero.

Dissipation potential \eqref{eq.Bingham.Xi} is not smooth at $\absnorm{\gradsym} = 0$, where it has no derivative. Therefore, Legendre transformation to $\Xi$ has to be generalized by means of subdifferentials, see e.g. \cite{Roubicek},
\begin{equation}
  \label{eq.LT.subdif}
  \dcstresssymb \in \partial \Xi,
\end{equation}
where $\partial \Xi$ is the set of slopes of all hyperplanes touching $\Xi$ at a point $\gradsym$. When constructing the subdifferential we assume that $\Xi$ is convex and continuous in a neighborhood of the point of non-smoothness. Continuity then guarantees that the set is non-empty (a consequence of Hahn--Banach theorem as commented in \cite{Roubicek}). Non-smoothness in non-convex regions can be handled as in Remark 5.8 in \cite{Roubicek}. When $\Xi$ is smooth, i.e. it has standard functional derivative, inclusion \eqref{eq.LT.subdif} becomes equality and the Legendre transformation restores its usual form. 

Inclusion \eqref{eq.LT.subdif} can be interpreted as the following problem: For a given $\dcstresssymb$ find all values of $\gradsym$ such that the inclusion is fulfilled. Solution to the problem is a graph $\gradsym(\dcstresssymb)$. It can be seen in Fig.~\ref{fig.Bingham.Xi} that for sufficiently high shear stresses, there is always a non-zero $\gradsym$ for which $\Xi_{\gradsym}=\dcstresssymb$. However, only $\partial \Xi$ at $\absnorm{\gradsym} = 0$ contains the low shear stresses, which means that when the shear stress is sufficiently low, only $\absnorm{\gradsym} = 0$ is the solution to the problem. In other words, there is no shear rate for sufficiently low shear stresses, but when shear stress crosses a threshold, shear rates starts to increase. We will return to this point graphically later on.

The conjugate dissipation potential is defined as
\begin{equation}
  \Xi^*(\dcstresssymb) = -\Xi \left( \gradsym(\dcstresssymb) \right) + \int\diff\rr\, \gradsym(\dcstresssymb):\dcstresssymb,
\end{equation}
and it is depicted in Fig.~\ref{fig.Bingham.Xic}. Since the conjugate dissipation potential is smooth, the backward Legendre transformation is carried out with derivatives as usually.

\begin{figure}[!htbp]
  \centering
  \includegraphics[width=0.6\textwidth]{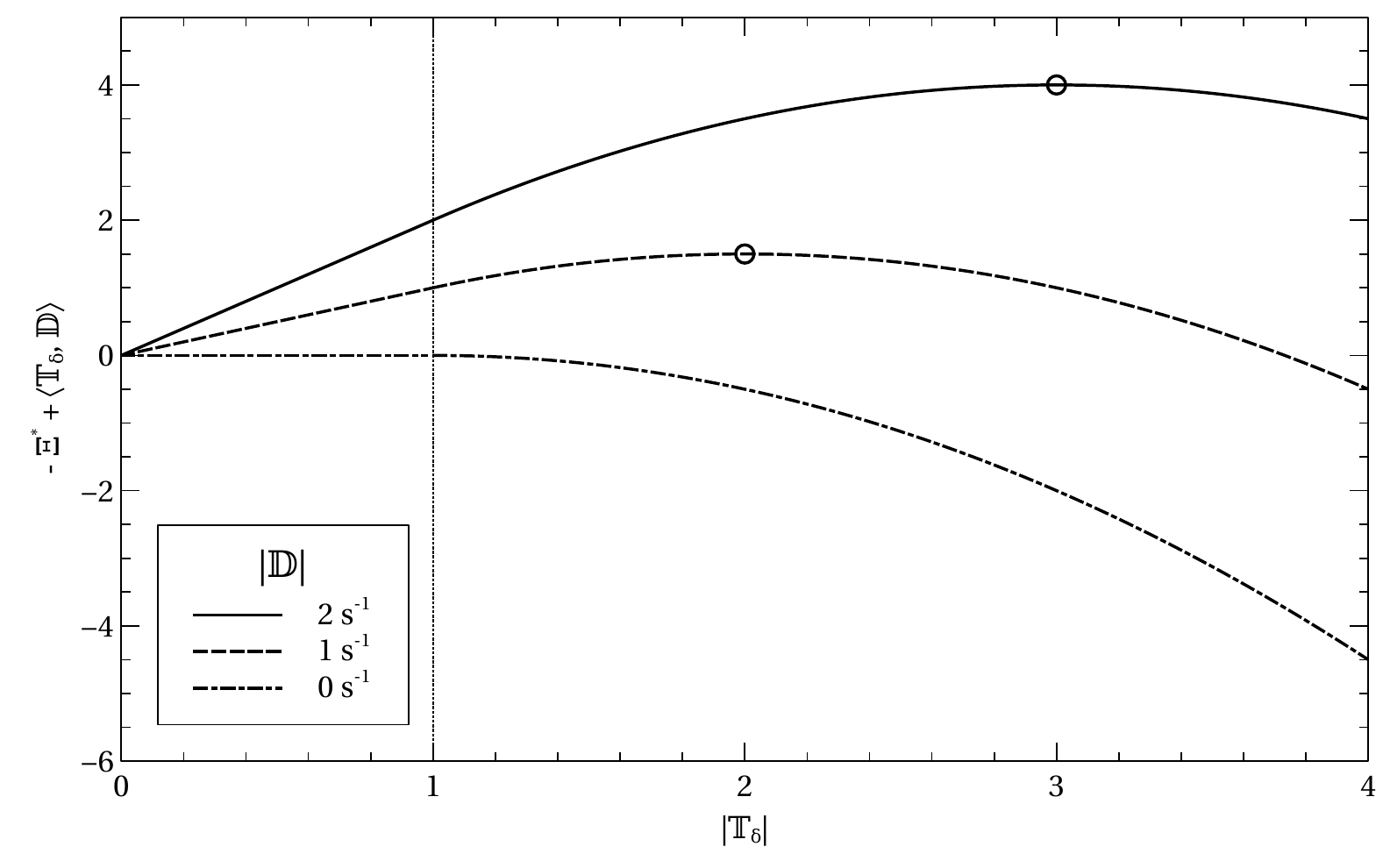}
  \caption{Reducing MTL function for the yield stress model with parameters $\alpha=\beta=1$. Shear stress is attracted to the local maxima. For $\absnorm{\gradsym} = 0$ the region near $\absnorm{\dcstresssymb} = 0$ is flat, which means that the shear stress can end up anywhere between $0$ and $1$. For higher shear rates a local maximum develops for non-zero stress.}
  \label{fig.Bingham.Massieu}
\end{figure}

Let us now have a look at these dissipation potentials from the perspective of multiscale mesoscopic thermodynamics as in Sec. \ref{sec.CR-thermodynamics}. The reducing MTL function is again $-\Xi^*(\dcstresssymb) + \int\diff\rr\, \dcstresssymb :\gradsym$ and it is plotted in Fig.~\ref{fig.Bingham.Massieu}. 
The shear stress again tends to the maximum of the MTL function for the given shear rate. When $\absnorm{\gradsym} = 0$, norm of shear stress can end up anywhere between 0 and 1 as the function is flat in that region. When shear rate is higher, the maximum is located to the right of $\absnorm{\dcstresssymb} = 1$. In other words, if there is non-zero shear rate, shear stress has to be higher than the threshold $\absnorm{\dcstresssymb} = 1$. The material thus exhibits yield stress behavior and its irreversible evolution is CR-stable.

  \section{Critical heat flux}
  \label{sec.CHF}
  
\begin{figure}[!htbp]
    \centering
    \includegraphics[width=0.6\textwidth]{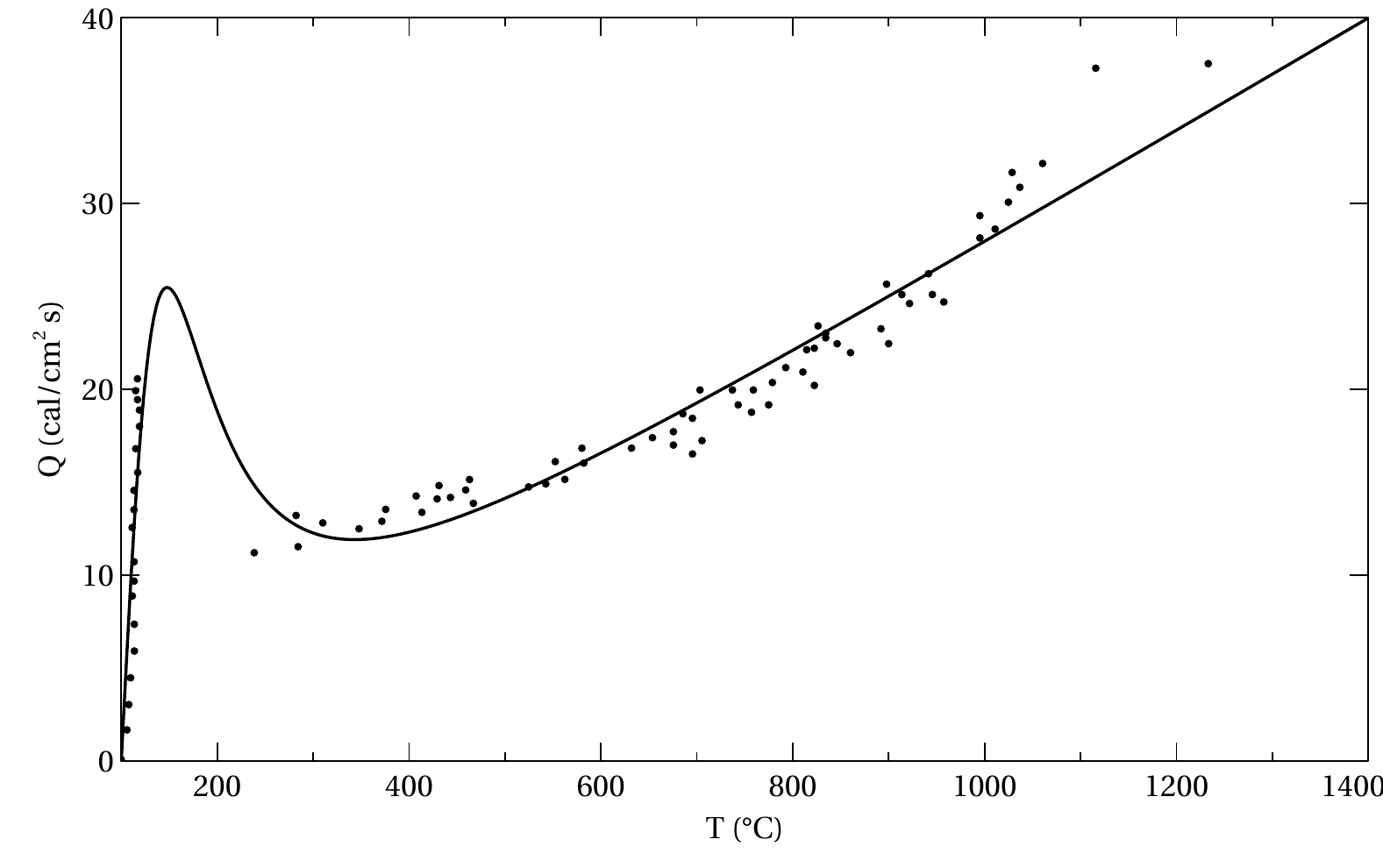}
    \caption{Boiling curve for a platinum wire of diameter $d = \unit[0.14]{mm}$ and boiling water $T = \unit[100]{^{\circ} C}$ from Nukiyama \cite{nukiyama.s:maximum}. The data were fitted with a relation of the form \eqref{eq:le-roux-rajagopal} with $\alpha = 0.98$, $\beta = 0.00026$, $\gamma = 0.03$ and $n = -1.41$.}
    \label{fig:boiling-curve}
  \end{figure}

  In 1934, Nukiyama \cite{nukiyama.s:maximum} experimentally showed that by controlling the heat flux from a hot platinum wire to boiling water, the temperature of the wire jumps between the heating branch and the cooling branch and exhibits a hysteresis loop. Then in 1937, Drew and Mueller \cite{drew.tb.mueller.ac:boiling} were able to control the temperature and observed that the heat flux initially grows with the temperature, then decreases until it finally grows again. The boiling curve, heat flux versus temperature difference (thermodynamic force) dependence, is then non-monotonous, see also Stosic \cite{Stosic} and the book by Bergman and co-authors \cite{bergman.tl:fundamentals}. In Fig.~\ref{fig:boiling-curve}, we have plotted the experimental data by Nukiyama \cite{nukiyama.s:maximum} and fitted them with a relation of the form \eqref{eq:le-roux-rajagopal}.

  As the flux-force relation can be described with an analogue of \eqref{eq:le-roux-rajagopal}, it can be likewise generated by a non-convex dissipation potential. Such a formulation would then lead to implications on the CR-stability of the constitutive relation and the experimentally observed hysteresis. For instance, the CR-unstable part is inaccessible when controlling the heat flux while careful temperature variation makes it available. This is analogous to the stress/shear rate behavior described in this paper. We hope to address this issue more explicitly in the near future.

\section{Conclusion}
\label{sec:conclusion}

Constitutive relation~\eqref{eq:le-roux-rajagopal}, that fits well the considered experimental data (Fig.~\ref{fig.exp}), can be formulated by means of dissipation potential $\Xi^*(\dcstresssymb)$, defined in Eq.~\eqref{eq.theta}. This dissipation potential is convex near equilibrium, but loses convexity for higher shear stresses. Due to the loss of convexity the Legendre-conjugate dissipation potential $\Xi(\gradsym)$ is multivalued. This reflects the experimental observation that varying shear stress leads to continuous evolution of shear rate (value of $\gradsym = \Xi^*_{\dcstresssymb}$ is unique) while varying shear rate leads to a jump (and possibly hysteresis) in shear stress (value of $\dcstresssymb = \Xi_{\gradsym}$ is not unique). The phenomenon of critical heat flux can be addressed analogically.

By extending classical hydrodynamics to extended hydrodynamics as in Sec.~\ref{sec.CR-thermodynamics}, one can show which parts of the multivalued dissipation potential $\Xi$ are CR-stable (with respect to perturbations of the constitutive relation), which are CR-metastable and which CR-unstable. It is then possible to identify between which branches of the dissipation potential the shear stress will jump, see Sec.~\ref{sec.stability}. The findings are compatible with the experimental data. 

\section*{Acknowledgment}
We are grateful to Vít Průša, Václav Klika and On\v{r}ej Sou\v{c}ek for their invaluable comments when discussing the manuscript and for pointing out further research directions (e.g. hysteresis modeling in rheology). We also want to express our gratitude to Gian Paolo Beretta for drawing our attention to the Critical heat flux.

Most ideas in this work are either inspired or brought up by Miroslav Grmela, founder of our thermodynamic tribe. 

This work was supported by Czech Science Foundation, project no.  17-15498Y. 
This work has been supported by Charles University Research program No. 
UNCE/SCI/023.

Adam Jane\v{c}ka acknowledges the support of Project No. LL1202 in the programme ERC-CZ funded by the Ministry of Education, Youth and Sports of the Czech Republic and the support of Project 260 449/2017 ``Student research in the field of physics didactics and mathematical and computer modelling''.

\bibliographystyle{spmpsci}       
\bibliography{library,implicit-cr}

\appendix

\section{Generalizations of  the Legendre transformation}
\label{sec.Legendres}
For a smooth, strictly convex (with positive second derivative) real function $f (x)$, the \emph{Legendre transform} (see \cite{Callen}) is defined as
\begin{equation}
  f^*_L (u) =_\bydefinition f(x) - u x,
\end{equation}
where $x$ satisfies $f'(x) = u$. Since the function $f$ is strictly convex, its derivative is monotonic and thus invertible, i.e. $x = \left( f' \right)^{-1} (u)$. Then we can write
\begin{equation}
  f^*_L (u) = f \left[ \left( f' \right)^{-1} (u) \right] - u \left( f' \right)^{-1} (u).
\end{equation}
The Legendre transform of a convex function is concave which can be immediately seen from
\begin{equation}
  \dd{f^*_L}{u} = \dd{f}{x} \dd{x}{u} - x - u \dd{x}{u} = -x, \quad \text{ and } \quad \ddd{f^*_L}{u} = - \dd{x}{u} = - \left( \ddd{f}{x}\right)^{-1}.
\end{equation}
Zia et al.~\cite{zia.rkp.ea:making} advocated for a definition with the opposite sign
\begin{equation}
  f^*_L (u) =_\bydefinition u x - f(x),
\end{equation}
which is a symmetric representation of the Legendre transform, $f^*_L(u) + f(x) = ux$, and converses convexity. Both versions are invertible and the bi-conjugate is the original function $f^{**} = f$. In the second version, the formula for the inverse transform is identical to the primal transform. 

The generalization of the Legendre transform to arbitrary functions is the \emph{Fenchel transform} (also called the Fenchel--Legendre transform or the Young--Fenchel conjugate)
\begin{equation}\label{eq.Fenchel}
  f^*_F (u) =_\bydefinition \sup_x \left[ u x - f(x) \right].
\end{equation}
The Fenchel transform is not invertible, the inversion of the conjugate yields the convex hull of the original function
\begin{equation}
  f^{**}_F (x) = \left( \overline{\conv} f \right) (x) \leq f(x),
\end{equation}
thus some information is lost in the process.

The Fenchel transform can be defined in the same manner for functionals, see \cite{bauschke.hh.lucet.y:what,Roubicek}. Let $V$ be a real Hilbert space and $f: V \to \R$, then the Fenchel transform $f^*_F$ at $u \in V^*$ is
\begin{equation}
  f^*_F (u) =_\bydefinition \sup_{v \in V} \left[ \langle u, v \rangle - f (v) \right],
\end{equation}
where $\langle \bullet, \bullet \rangle$ is the corresponding inner product. The conjugate $f^*_F$ is always convex and lower semicontinuous and $f^{**}_F = f$ if and only if $f$ is convex and lower semicontinuous.

For arbitrary differentiable functions, \cite{dorst.l.boomgaard.r:analytical,dorst.l.boomgaard.r:morphological} introduced the invertible multivalued \emph{slope transform}
\begin{equation}
  f^*_S (u) =_\bydefinition \stat_x \left[ u x - f(x) \right],
\end{equation}
where $\stat_x f(x)$ is the set of all stationary values of $f(x)$ with respect to $x$
\begin{equation}
  \stat_x f(x) =_\bydefinition \left\{ f(x): f'(x) = 0 \right\}.
\end{equation}
Then the inverse slope transform is
\begin{equation}
  f(x) =_\bydefinition \stat_u \left[ u x - f^*_S (u) \right].
\end{equation}
Throughout the paper, we use this general definition of the Legendre-like conjugation for it is invertible and no information is lost. For functions with consecutively variating convex and concave parts, each with an invertible derivative (as in our case), the slope transform corresponds to application of the classical Legendre transform to each part. Then, the number of different functions in the multivalued Legendre transform agrees to the convex/concave parts of the original function, see~\cite{maragos.p:slope}. For functionals in general, we believe that the slope transform should be invertible as well, unfortunately we could not find any relevant mathematical reference.


  \section{Reversibility and irreversibility}
  \label{sec.rev}
  
  Definition of reversible dynamics is based on the time-reversal transformation (TRT), see e.g. \cite{PRE2014} or \cite{HCO}. TRT inverts momenta of all particles, $\pp\stackrel{TRT}{\rightarrow}-\pp$ and the direction of time increment $\diff t\stackrel{TRT}{\rightarrow} -\diff t'$. If an evolution equation is transformed by TRT to itself, it is reversible. If, on the other hand, some terms on the right hand side transform differently than the left hand side (time derivative of a state variable), these terms generate irreversible evolution.

  Let us illustrate the concept of reversibility on the Hamilton canonical equations, which are transformed by TRT as follows.
  \begin{subequations}
    \begin{align}
      \dd{\rr}{t} &= H_\pp \stackrel{TRT}{\rightarrow} - \dd{\rr}{t'} = -H_{\pp}, \\
      \dd{\pp}{t} &= -H_\rr \stackrel{TRT}{\rightarrow} \dd{\pp}{t'} = -H_{\rr}.
    \end{align}
  \end{subequations}
  After TRT, the Hamilton canonical equations are equivalent with the original equations, which means that they are reversible. 

  On the other hand, Hamilton canonical equations with damping, e.g. 
  \begin{subequations}
    \begin{align}
      \dd{\rr}{t} &= H_\pp \stackrel{TRT}{\rightarrow} - \dd{\rr}{t'} = -H_{\pp}, \\
      \dd{\pp}{t} &= -H_\rr - \alpha \pp \stackrel{TRT}{\rightarrow} \dd{\pp}{t'} = -H_{\rr} + \alpha\pp,
    \end{align}
  \end{subequations}
  are not completely reversible because the damping term in the momentum equation transforms differently than the left hand side of the equation. The equations consist of both reversible and irreversible evolution.

  Note also that the equations obtained from a variational principle (extremization of an action) are not automatically reversible although Hamilton canonical equations (without damping) are. Such equations can be found for instance in textbook \cite{Landau5}, where a dissipation function is introduced into the principle of least action.

  On the other hand, Hamiltonian kinematics derived by projection from a more detailed reversible Hamiltonian kinematics is surely reversible, see \cite{PhysicaD-2015}, and this is the case the Boltzmann kinematics (kinetic theory) and fluid mechanics. When talking about reversible and irreversible terms in this paper, we mean the reversibility with respect to TRT and we refer to the cited literature in case of any doubts.

\section{Reversible part of momentum balance}
\label{sec.hydro}

The reversible part of the momentum balance (evolution equation for momentum density) is generated by the Poisson bracket of classical hydrodynamics (see e.g. \cite{Arnold}), 
\begin{multline}
  \{A,B\} = \int \diff\rr\, \rho\left(\partial_i A_\rho B_{u_i} - \partial_i B_\rho A_{u_i}\right) \\
  + \int \diff\rr\, u_i\left(\partial_j A_{u_i} B_{u_j} - \partial_j B_{u_i}A_{u_j}\right) \\
  + \int \diff\rr\, s\left(\partial_i A_s B_{u_i} - \partial_i B_s A_{u_i}\right).
\end{multline}
Supplied with energy, e.g.
\begin{equation}
  \label{eq.E.hydro}
  E = \int \diff\rr\, \frac{\uu^2}{2\rho} + \eps(\rho,s),
\end{equation}
the reversible evolution of an arbitrary functional of the hydrodynamic variables $A(\rho, \uu, s)$ is 
\begin{equation}\label{eq.Adot}
\pd{A}{t} = \{A,E\}.
\end{equation}
By rewriting time-derivative of the functional as
\begin{equation}
  \pd{A}{t} = \int \diff\rr\, A_\rho \pd{\rho}{t} + A_{u_i} \pd{u_i}{t} + A_s \pd{s}{t},
\end{equation}
and comparing with Eq.~\eqref{eq.Adot}, one can read the reversible part of evolution equations for the hydrodynamic fields, 
\begin{subequations}
  \begin{eqnarray}
    \pd{\rho}{t} &=& -\partial_i (\rho E_{u_i})=-\partial_i u_i, \\
    \label{eq.u.evo.rev} \pd{u_i}{t} &=& -\partial_j \left(u_i E_{u_j} \right) -\rho \partial_i E_\rho - u_j \partial_i E_{u_j} -s \partial_i E_s = -\partial_j \left(\frac{u_i u_j}{\rho}\right) - \partial_i p, \\
    \pd{s}{t} &=& -\partial_i (s E_{u_i})= -\partial_i \left(\frac{s u_i}{\rho}\right),
  \end{eqnarray}
\end{subequations}
where $p=-\eps(\rho,s) + \rho \eps_\rho + s \eps_s$ is the pressure. The reversible evolution of hydrodynamic fields with energy \eqref{eq.E.hydro} is of course the Euler equations. The reversible terms from Eq.~\eqref{eq.u.evo} are on the right hand side of Eq.~\eqref{eq.u.evo.rev}.

Strictly speaking, Eq.~\eqref{eq.u.evo} is in the entropic representation (in variables $(\rho,\uu,e)$) while Eq.~\eqref{eq.u.evo.rev} is in the energetic representation (variables $(\rho,\uu,s)$) and one should thus perform transformation from the latter representation to the former, see e.g. \cite{HCO}.

\section{Kinetic theory}
\label{sec.KT}

Let us recall a few fundamental findings from classical kinetic theory, well reviewed in \cite{dGM}. The state variable of kinetic theory is the one-particle distribution function $f(\rr,\vv)$, which depends on the position and the velocity. To express energy of interactions between particles exactly, one would need to use a two-particle distribution function, which is inaccessible in the one-particle kinetic theory. Therefore, let us stay in the region of ideal or nearly ideal (diluted) gas, where total energy is
\begin{equation}
  E = \int \diff \rr\, \int \diff \vv\, \frac{1}{2} m \vv^2 f(\rr,\vv),
\end{equation}
$m$ being mass of one particle. Density, momentum density and entropy density then are
\begin{subequations}
\begin{eqnarray}
  \rho(\rr) &=& \int \diff \vv\, m f(\rr,\vv), \\
  \uu(\rr) &=& \int \diff \vv\, m \vv f(\rr,\vv),\\
  s(\rr) &=& -k_B\int \diff \vv\,  f(\rr,\vv)\ln(f(\rr,\vv)-1),
\end{eqnarray}
where Boltzmann entropy was chosen to be the entropy of the system.
\end{subequations}

Evolution equation for the one-particle distribution function in absence of external force is the Boltzmann equation
\begin{equation}
  \pd{f}{t} = -\vv \cdot \nabla_{\rr} f + \left( \pd{f}{t} \right)_{irr},
\end{equation}
where the last term represents the collision integral, which generates irreversible evolution of $f$.
Evolution equation for momentum density then becomes
\begin{equation}
  \pd{\uu}{t} = \divergence \left( -\int \diff \vv\, m \vv\otimes\vv f(\rr,\vv) \right).
\end{equation}
Note that the collision integral does not contribute to balance of momentum due to conservation of momentum in collisions. The term on the right-hand side generates reversible evolution of $\uu$ on the level of kinetic theory, since it is even with respect to time-reversal, see e.g. \cite{PRE2014}, and it can be regarded as divergence of stress tensor
\begin{equation}
  \label{eq.KT.T}
  \tensorq{T} = -p \identity + \frac{\uu\otimes\uu}{\rho} + \underbrace{\int\diff\vv\, m \left(\vv-\frac{\uu}{\rho}\right)\otimes\left(\vv-\frac{\uu}{\rho}\right) f}_{=\dcstresssymb}.
\end{equation}
The last term in Eq.~\eqref{eq.KT.T} constitutes the shear stress, and the standard expression for Newtonian fluids can be recovered near equilibrium for example by means of Champan--Enskog expansion, see e.g. \cite{dGM,MG-CR}. The important observation here is that on the kinetic level of description the shear stress (or extra stress) is even with respect to time-reversal while on the hydrodynamic level, where it is proportional to shear rate, it is odd. 

\end{document}